\documentclass[sigconf]{acmart}

\AtBeginDocument{%
  \providecommand\BibTeX{{%
    \normalfont B\kern-0.5em{\scshape i\kern-0.25em b}\kern-0.8em\TeX}}}

\copyrightyear{2021}
\acmYear{2021}
\setcopyright{acmcopyright}\acmConference[SEC '21]{The Sixth ACM/IEEE Symposium on Edge Computing}{December 14--17, 2021}{San Jose, CA, USA}
\acmBooktitle{The Sixth ACM/IEEE Symposium on Edge Computing (SEC '21), December 14--17, 2021, San Jose, CA, USA}
\acmPrice{15.00}
\acmDOI{10.1145/3453142.3491279}
\acmISBN{978-1-4503-8390-5/21/12}

\usepackage{microtype}
\usepackage{graphicx}
\usepackage{caption}
\usepackage{subcaption}
\usepackage{booktabs} 
\usepackage{tabularx}
\usepackage{multicol}
\usepackage{empheq}
\usepackage{amsmath}
\usepackage{todonotes}
\usepackage{xspace}

\usepackage{hyperref}

\newcommand{\figref}[1]{Figure \ref{#1}}
\newcommand{\secref}[1]{Section \ref{#1}}
\newcommand{\tabref}[1]{Table \ref{#1}}
\newcommand{\approach}{AQuA}

\begin{document}

\title{\approach: \underline{A}nalytical \underline{Qu}ality \underline{A}ssessment for Optimizing Video Analytics Systems}

\author{Sibendu Paul}
\affiliation{%
  \institution{Purdue University}
  \city{West Lafayette}
  \country{USA}
}
\email{paul90@purdue.edu}

\authornote{Work mostly done as an intern at NEC Laboratories America.}

\author{Utsav Drolia}
\affiliation{%
  \institution{NEC Laboratories America}
  \city{San Jose}
  \country{USA}
}
\email{utsavdrolia@gmail.com}

\authornote{Work done when Utsav Drolia was a Researcher at NEC Laboratories America.}

\author{Y. Charlie Hu}
\affiliation{%
  \institution{Purdue University}
  \city{West Lafayette}
  \country{USA}
}
\email{ychu@purdue.edu}

\author{Srimat T. Chakradhar}
\affiliation{%
  \institution{NEC Laboratories America}
   \city{New Jersey}
  \country{USA}
}
\email{chak@nec-labs.com}



\renewcommand{\shortauthors}{Sibendu Paul, et al.}

\begin{abstract}
Millions of cameras at edge are being deployed to power a variety of different deep learning applications.
However, the frames captured by these cameras are not always pristine - they can be distorted due to lighting issues, sensor noise, compression etc. Such distortions not only deteriorate visual quality, they impact the accuracy of deep learning applications that process such video streams.
In this work, we introduce \approach, to protect application accuracy against such distorted frames by  scoring the level of distortion in the frames. It takes into account the analytical quality of frames, not the visual quality, by learning a novel metric, \emph{classifier opinion score}, and uses a lightweight, CNN-based, object-independent feature extractor.
\approach\ accurately scores distortion levels of frames and generalizes to multiple different deep learning applications. When used for filtering poor quality frames at edge, it reduces high-confidence errors for analytics applications by 17\%. Through filtering, and due to its low overhead (14ms), \approach\ can also reduce computation time and average bandwidth usage by 25\%.
\end{abstract}

\maketitle

\section{Introduction}
\label{section:intro}

\begin{figure}[b]
    \centering
    \includegraphics[width=\columnwidth]{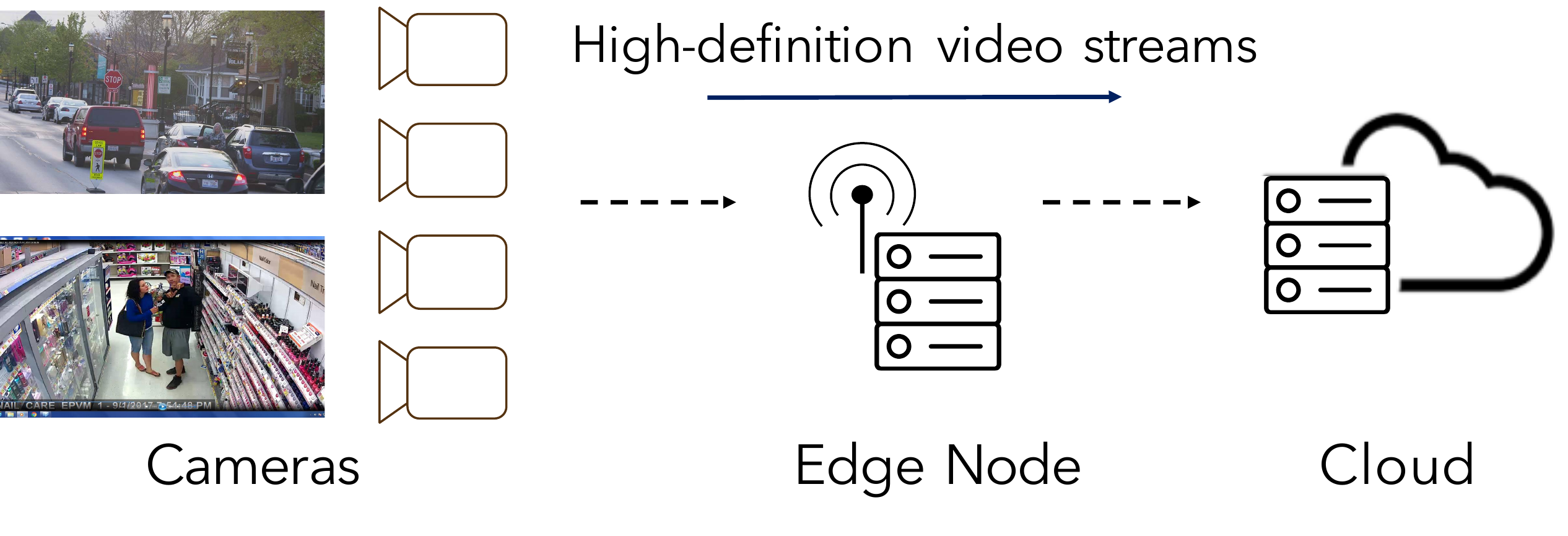}
    \caption{Large-scale, real-time video analytics deployment}
    \label{fig:app_introduction}
\end{figure}

Video camera deployments are increasing rapidly, powering applications like city-scale traffic analytics, security and retail analytics. A recent report estimated the market size of video analytics to be \$4.10 billion in 2020, and   \$20.80 billion by 2027~\cite{VIDEO-ANALYTICS-MARKET}. A CNBC study reported that by 2021, about  one billion surveillance cameras will be ensuring our safety and security~\cite{CNBC-STUDY}. \figref{fig:app_introduction} illustrates how such cameras can continuously capture high-resolution video of the real world and transmit it to application services running on nearby edge computing nodes or on the cloud.
The exponential growth of camera deployments and video analytics applications can be attributed to two main reasons - deep learning, which is enabling accurate computer vision applications~\cite{krizhevsky2012imagenet}, and 5G , which is making low-latency and high-bandwidth communication possible~\cite{CNET-5G,QUALCOMM-5G}.

Several factors impact the quality of data acquired by the cameras. In-camera distortions are introduced by camera hardware or on-board software during the video capture process. Such distortions include texture distortions, artifacts due to exposure and lens limitations, focus and color aberrations. Factors such as lighting (low-light, glare and haze), noise sensitivity, acquisition speed, camera setup, and camera shake can also adversely affect a video’s perceived visual quality.  Some distortions, like exposure and color-related distortions, for instance, occur more frequently than others.
Different forms of distortion can also be introduced 
during post-capture. For example, video compression (like H.264, MPEG or VP9 encoding) is lossy, and transmission over IP networks or wireless networks is error-prone, and they can introduce distortions that adversely affect the perceived video quality. Note that all these distortions occur naturally in video acquisition and transmission,  and they are  not introduced by an adversary.
Irrespective of the source of distortions, both content and network providers are deeply invested in finding better ways to monitor and control video quality. Designing reliable predictive models and algorithms for detecting, and ameliorating distortions is of great interest \cite{chak-video-distortions1, chak-video-distortions2}.
\begin{figure}
    \centering
    \begin{subfigure}[b]{0.3\columnwidth}
        \centering
        \includegraphics[width=\textwidth]{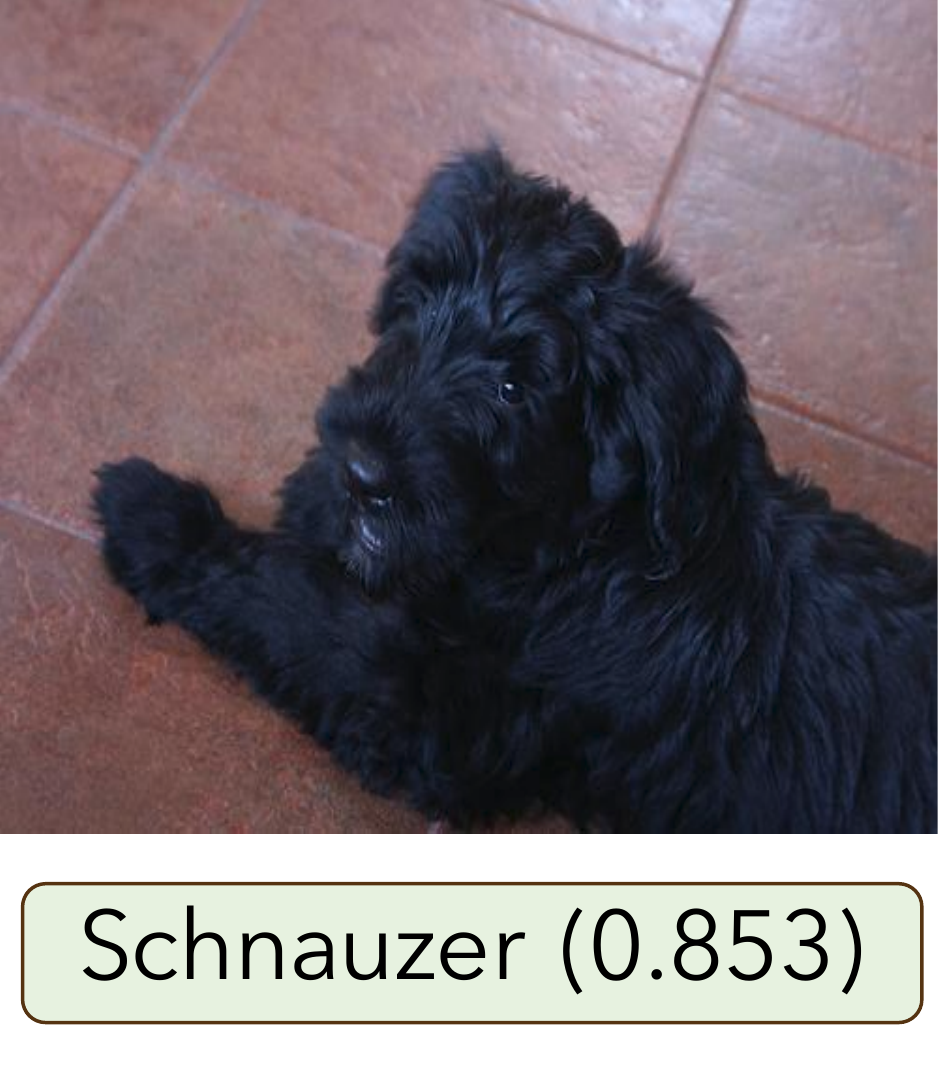}
        \caption{Original}
        \label{fig:intro_orig}
    \end{subfigure}
    \begin{subfigure}[b]{0.3\columnwidth}
        \centering
        \includegraphics[width=\textwidth]{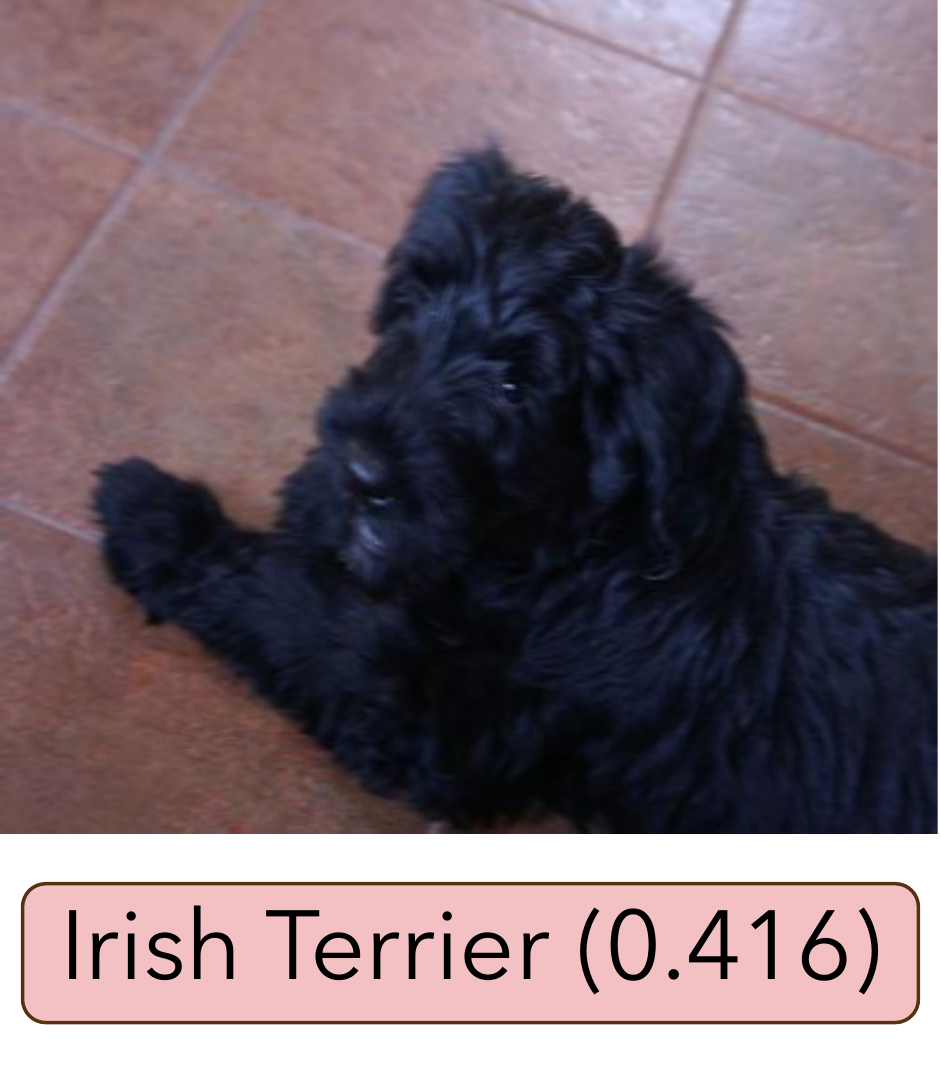}
        \caption{Blurred}
        \label{fig:intro_blur}
    \end{subfigure}
    \begin{subfigure}[b]{0.3\columnwidth}
        \centering
        \includegraphics[width=\textwidth]{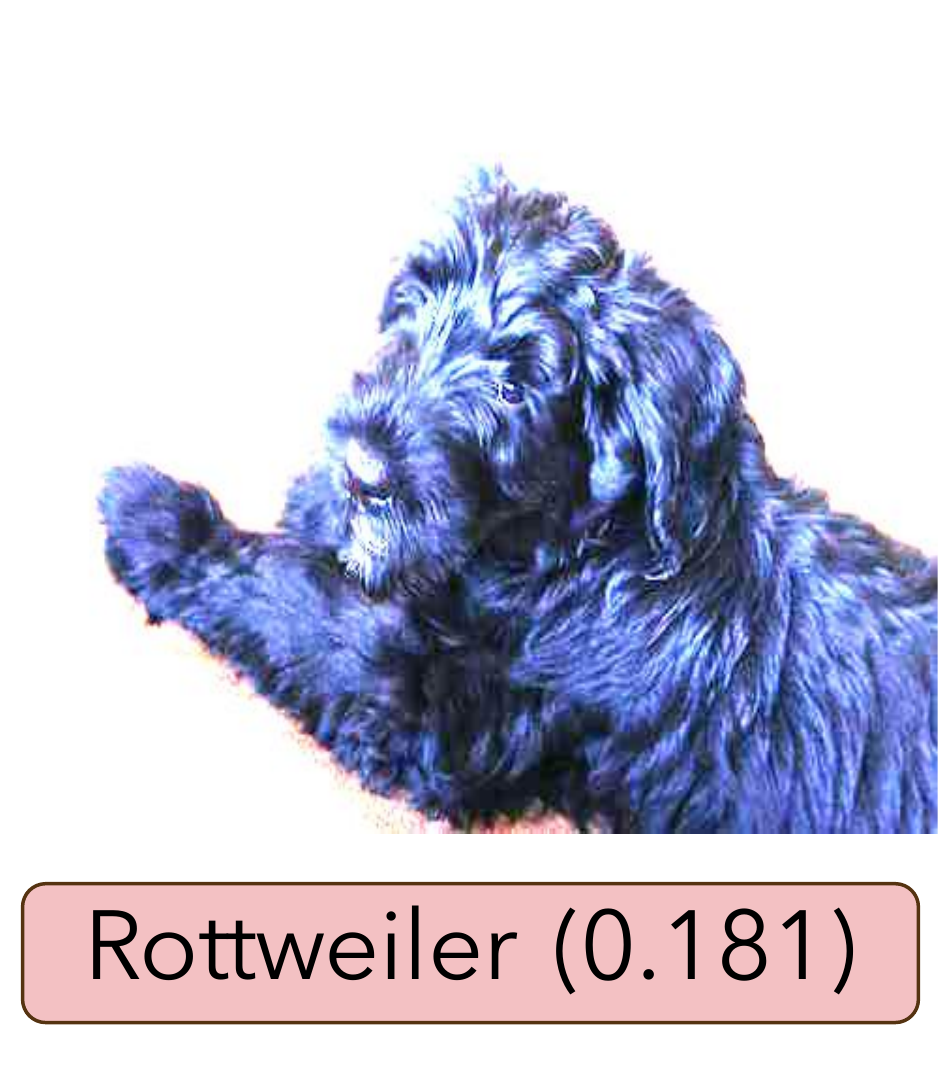}
        \caption{Over-exposed}
        \label{fig:intro_over_exposed}
    \end{subfigure}
    \caption{Impact of distortions on accuracy}
    \label{fig:intro_accuracy_impact}
\vspace{-0.2in}
\end{figure}

\begin{figure*}
     \centering
     \includegraphics[width=\textwidth]{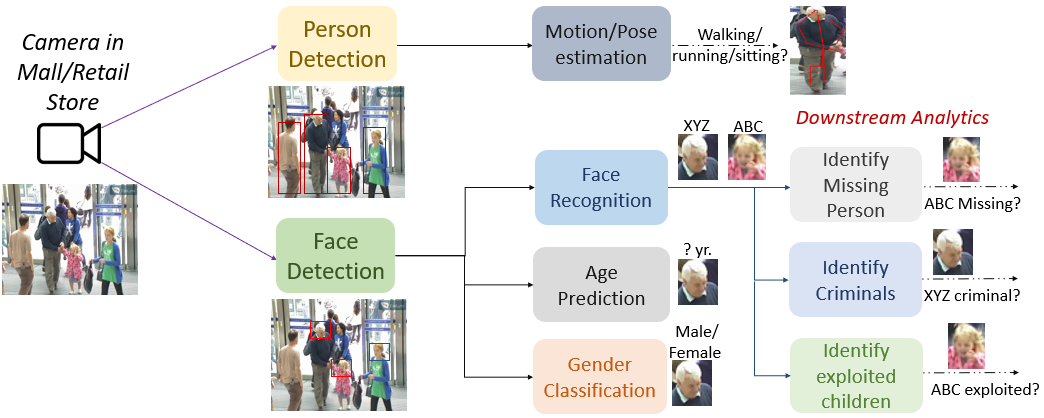}
     \caption{Multiple Analytics performed on Same Video Stream \emph{(Mall)}}
 \label{fig:app_mul_analytics}
 \end{figure*}

\figref{fig:intro_accuracy_impact} shows a few examples of distortions (for more examples of distortions, please see Figure 2 in \cite{chak-video-distortions1}).
These distortions do lower the perceived image quality (as observed by a human being), but more importantly, they adversely affect the accuracy of video analytics applications.
\figref{fig:intro_accuracy_impact} shows examples of adverse effects that distorted images can have on a state-of-the-art image classifier, ImageNet-trained ResNet101\footnote{We chose ResNet because recent computer vision models rely on DCNNs and use such image-classifiers as the base network.}. Although the original version of the image is correctly classified, slight distortions of the image result in mis-classifications. For example,  \figref{fig:intro_blur} has minuscule motion blur, but the classifier is {\it confidently} wrong (there are a 1000 classes, and the classifier has a 41.6\% confidence in its mis-prediction). These \textit{high-confidence errors} adversely impact the accuracy of the application and they cannot be filtered out from further consideration by using simple thresholds on prediction confidence. 
In \figref{fig:intro_over_exposed}, although the classifier again predicts incorrectly, it is not confident in its prediction. This can be filtered out by using suitable thresholds on prediction confidence. However, the distorted image had to be transmitted from the camera to the application service, processed by the computationally-expensive deep learning classifier, only to be filtered out due to low confidence. 
In a typical edge-assisted video analytics system, there are multiple analytics applications that are simultaneously analyzing a video stream 
(as shown in ~\figref{fig:app_mul_analytics}), 
and  negative consequences of wasted computations or high-confidence errors due to low quality input frames do snowball.

In this paper, our goal is to detect and 
score poor quality frames from video streams as early as possible, ideally immediately after capture at edge (either edge node or edge camera in ~\figref{fig:app_introduction}), and enable other actions so that overall accuracy of the applications increase, while compute and network resource usage decreases. In the present work, we filter these poor quality frames. In our on-going work, we are exploring other actions like alerting operators about the quality, establishing a feedback loop with the camera to dynamically adapt its settings to improve quality, marking frames as prospects for future fine-tuning, etc. 

At first glance, the obvious approach to detect these low quality frames is to use state-of-the-art image quality assessment (IQA) tools~\cite{biqi_moorthy2010two, divine_moorthy2011blind, blinds2_saad2012blind, gmlog_xue2014blind, kang2014convolutional,jin2018ilgnet} that score the perceptual quality of an image. However, as we show later (\secref{subsection:iqa_correlation}), we observed that these IQA models'~\cite{brisque_mittal2012no,talebi2018nima} image quality assessments do not align with a classifier's assessment of the ``quality", as evidenced by the classifier's confidence in the correct class.
Motivated by this finding, and inspired by the human opinion score used by IQA models, we introduce the notion of \textit{classifier opinion score}, which captures the classifier's assessment of the image quality. Armed with the classifier's opinion score, we present~\approach, an analytical quality assessor for images. Our approach judges if a frame is good enough for further analytics or not, and assigns a quality score accordingly. We also construct a filtering system based on~\approach\ that can identify, flag and/or discard distorted frames immediately after capture, or after video compression and transmission.

In this paper, we make the following contributions:
\begin{enumerate}
    \item We show empirically that image quality assessments from state-of-the-art IQA models do not align with a classifier's assessment of image quality, thereby leading to mis-classification and  high-confidence errors. 
    \item We propose two new metrics, \emph{Mean Classifier Opinion Score} (MCOS), and its semi-supervised version MCOS$_{SS}$, which are used for training a new, deep-learning based analytical quality assessor. To our knowledge, this is the {\it first} time that a classifier's notion of image quality has been defined, quantified, and used to train an effective image quality assessment model.
    \item We design and train a new, \emph{lightweight} feature extractor, which leverages early layers of pre-trained image classifiers to quickly learn low-level image features. Our model is \emph{10x} faster than state-of-the-art image classifiers, which makes our model to be a good fit for resource-constrained mobile, embedded or edge processing environments.
    \item We implement~\approach, a deep-learning model that leverages classifier opinion scores to estimate a frame's analytical quality. To our knowledge, this is the {\it first} system to explicitly consider a classifier's assessment of image quality, and thus improve any real-time video analytics pipeline.
    \item We conduct multiple evaluations of~\approach\ to show its accuracy, efficacy as a filter, and quantify its impact on application accuracy and resource usage.
\end{enumerate}

\approach\ enables filtering frames with high precision and recall compared to existing IQA models. It can be used for multiple computer vision applications such as object detection, instance segmentation, and pose estimation.
When evaluated on a real-world application (face recognition), it can reduce false positives by 17\% (3x more than BRISQUE). By filtering low quality frames, coupled with its low-overhead of only 14ms,~\approach\ reduces computation and communication resource requirements for both edge-only and edge-cloud systems. 

\begin{figure*}[h]
\begin{tabularx}{\linewidth}{@{}cXX@{}cXX@{}cXX@{}cXX@{}}
\begin{tabular}{ccccc}
    \subfloat[Original image]{\includegraphics[width=3.2cm]{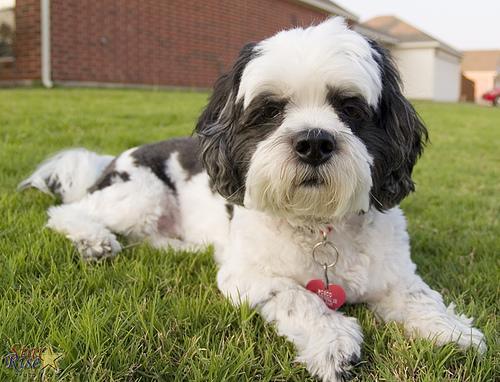}} 
    & \subfloat[Over exposed]{\includegraphics[width=3.2cm]{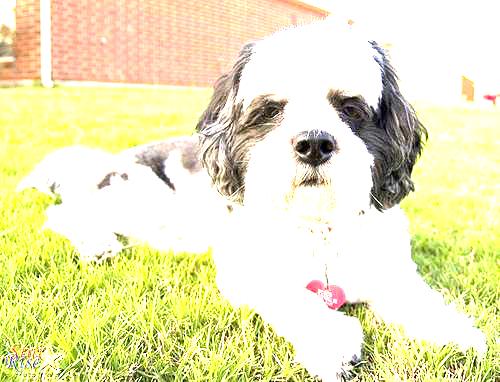}}
    & \subfloat[Under exposed]{\includegraphics[width=3.2cm]{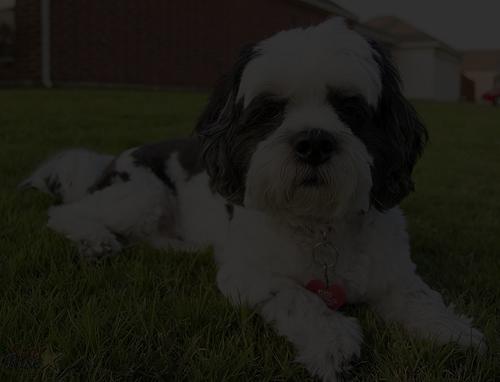}}
    & \subfloat[Low Contrast]{\includegraphics[width=3.2cm]{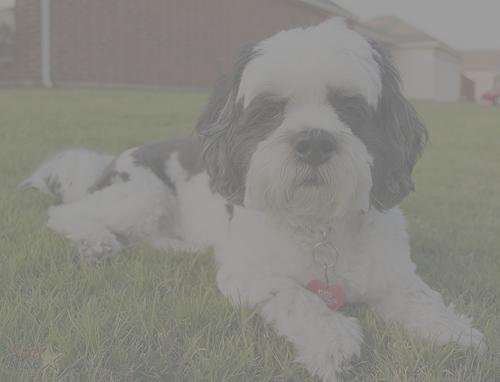}}
    & \subfloat[High Contrast]{\includegraphics[width=3.2cm]{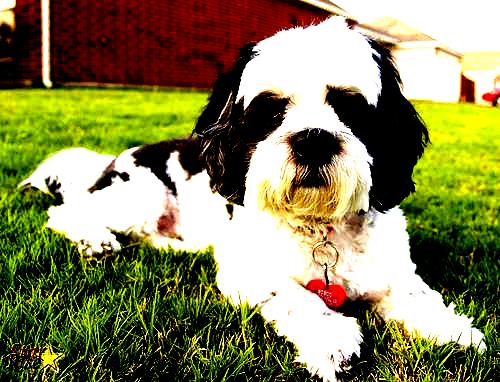}}\\
    \subfloat[Motion blur]{\includegraphics[width=3.2cm]{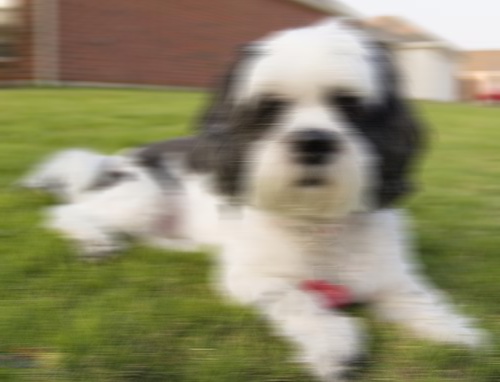}}
    & \subfloat[Compression Artifact]{\includegraphics[width=3.2cm]{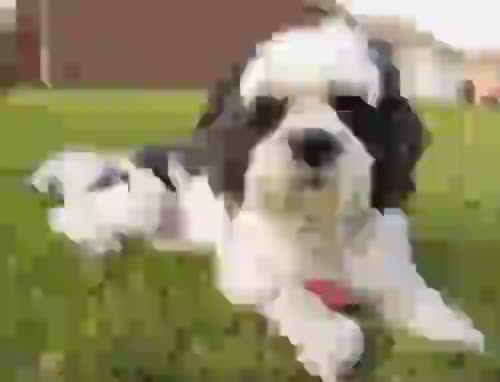}}  
    & \subfloat[Low-light noise]{\includegraphics[width=3.2cm]{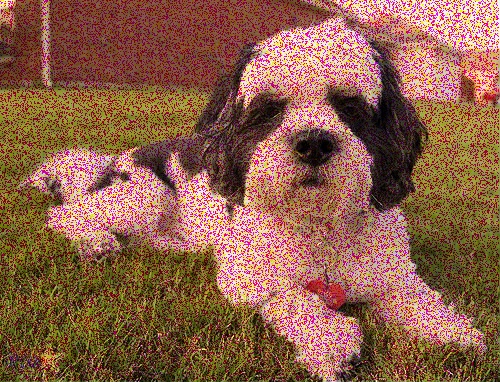}}
    & \subfloat[Defocus Blur]{\includegraphics[width=3.2cm]{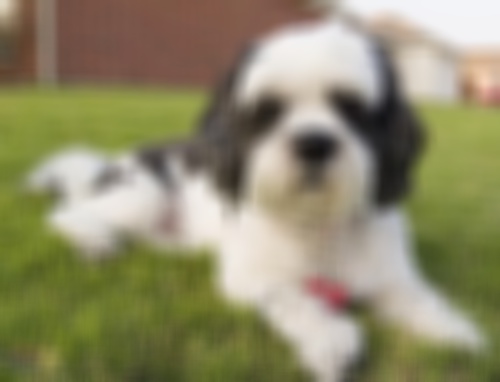}}
    & \subfloat[Gaussian Noise]{\includegraphics[width=3.2cm]{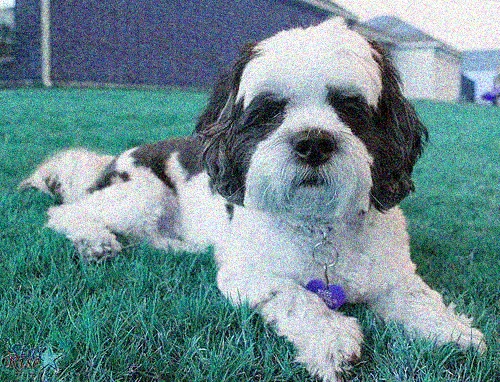}
    }\\
\end{tabular}
\end{tabularx}
\caption{Types of distortions and their visual impact.}
\label{fig:diff_distortions}
\end{figure*}

\section{Background}
\label{section:background}
\subsection{Types of visual distortions}

Visual distortions manifest as noise, artifacts or loss of detail in a frame. These distortions can occur due to multiple factors and are grouped under two broad categories depending on when the distortion occurs, (1) Image Acquisition and (2) Image Transmission. Under image acquisition, distortions can happen due to incorrect settings of the camera, such as focus (focal blur), exposure settings (over or under exposure) or shutter speed (motion blur). Cameras using a low quality sensor may add Gaussian noise~\cite{dodge2016understanding} to the frame, or cause sparse but intense disturbances at low-light. 

For efficient image transmission, raw frames need to undergo compression, such as H.264, MJPEG and HEVC. These compression algorithms are typically lossy and can induce artifacts like blocking and blurring. 

The effects of such distortions can be seen in \figref{fig:diff_distortions}.

\subsection{Image Quality Assessment}
\label{subsection:iqa}
\begin{figure}[t]
    \centering
    \includegraphics[width=\columnwidth]{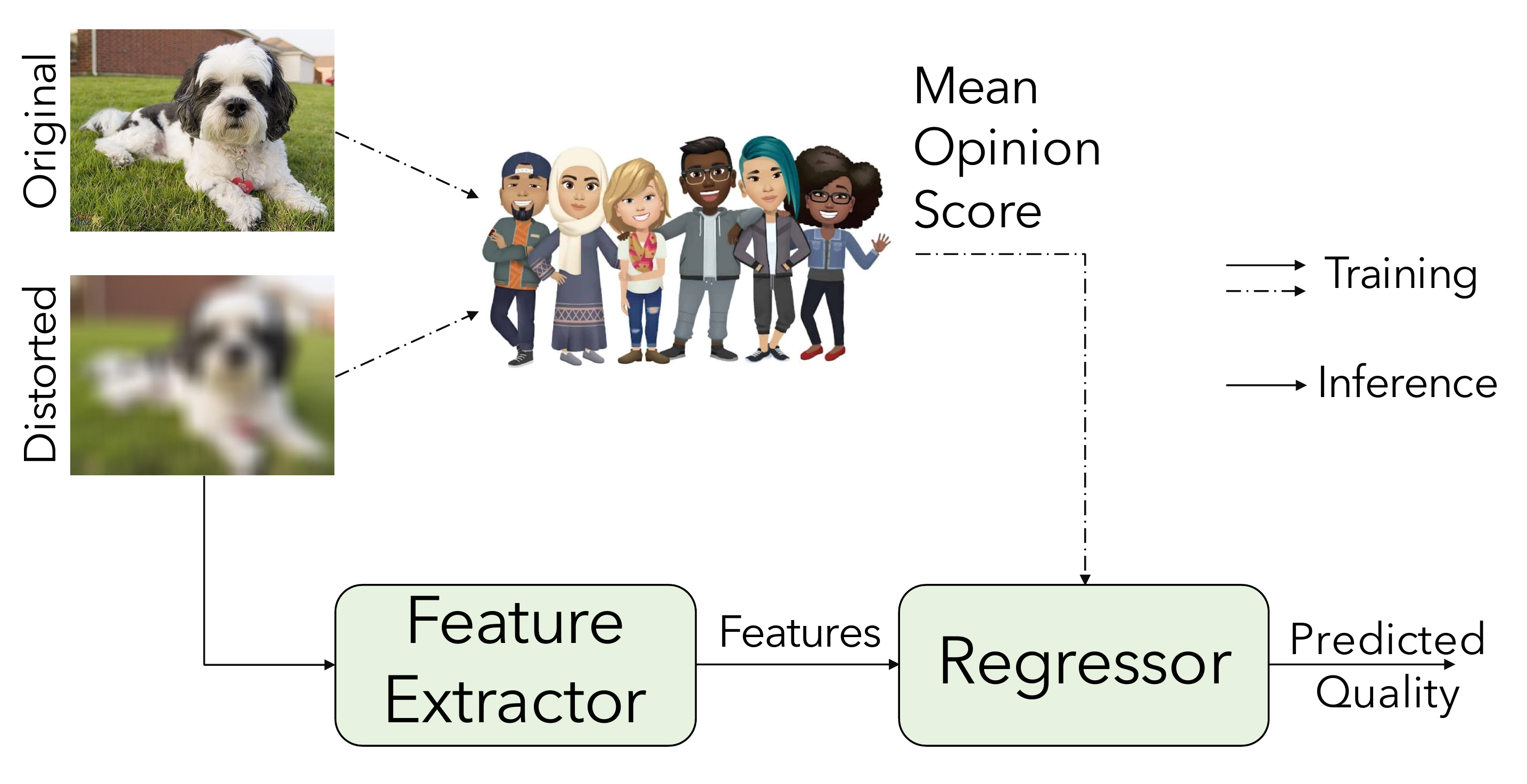}
    \caption{Training a typical IQA model}
    \label{fig:iqa_training}
    \vspace{-0.2in}
\end{figure}

Image Quality Assessment (IQA) techniques are used to score the visual quality of images. They typically take the form of a machine learning model, which is trained to estimate a human observer's opinion of a given input image.
The training data for these models includes original and distorted images (X), and the human opinion score for each image (Y). Human observers rate the difference between the original image and its distorted version as a score. A large opinion score implies higher level of distortion present in the image under consideration. TID2013~\cite{TID2013}, AVA~\cite{AVA}, LIVE~\cite{LIVE} are a few commonly used training datasets.
These models employ a two-stage framework: feature extraction followed by regression. 
Figure~\ref{fig:iqa_training} shows how a typical IQA model is trained and used. 

Early IQA algorithms used natural scene statistics (NSS) based feature extraction, which encodes the resultant distribution from image filters. These included BIQI~\cite{biqi_moorthy2010two}, DIIVINE~\cite{divine_moorthy2011blind}, BLINDS-II~\cite{blinds2_saad2012blind}, GMLOG~\cite{gmlog_xue2014blind} and BRISQUE~\cite{brisque_mittal2012no}. 

CORNIA~\cite{cornia_ye2012unsupervised} was the first to propose that image features can be learnt directly from raw pixels using CNNs. CORNIA's success motivated other CNN-based algorithms such as ~\cite{kang2014convolutional}, ILG-net ~\cite{jin2018ilgnet} and Neural Image Assessment (NIMA) ~\cite{talebi2018nima}. 

\subsection{Image Classification}
Image classification is regarded as a basic computer vision task, where the input image is classified according to what object(s) it contains. With the advent of convolutional neural networks~\cite{lecun2015lenet} combined with deep learning~\cite{krizhevsky2012imagenet}, models have now started to surpass human-level accuracy~\cite{vgg19simonyan2014very} for image classification on large datasets~\cite{deng2009imagenet}.
These high-accuracy classifiers are also used as backbones for other computer vision tasks such as object detection~\cite{ren2015faster, yoloredmon2016you}, pose-estimation~\cite{moon2019posefix}, instance segmentation~\cite{qiao2020detectors}. Thus, we use image classification as a running example for a basic  unit that any video analytics application might have.

\section{Motivation}
\label{section:motivation}
We first show  that distortions not only adversely affect the perceptual  quality of frames (as observed in~\figref{fig:diff_distortions}), but they also cause classifiers to make errors. Then, we show that traditional image quality assessment, based on perceptual quality, is not up to the task of identifying distorted images on which a classifier would falter.

\subsection{Adverse effect of distortions on classifier accuracy}
\label{subsection:distortions_accuracy}
\begin{figure}

\def\tabularxcolumn#1{m{#1}}
\begin{tabularx}{\linewidth}{@{}cXX@{}}
\begin{tabular}{cc}
\subfloat[Brightness]{\raggedleft\includegraphics[width=4cm]{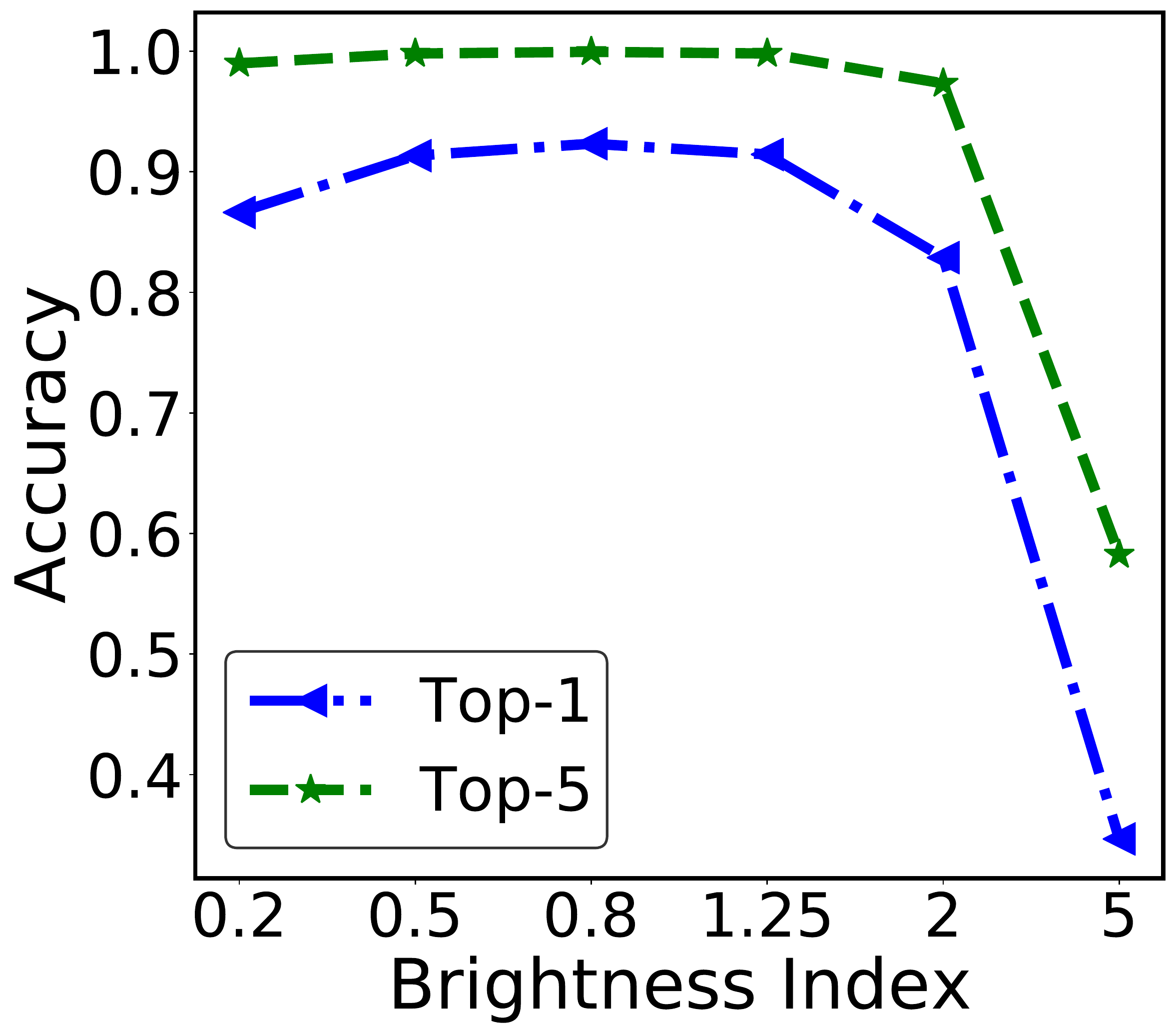}}
   & \subfloat[Motion Blur]{\raggedright\includegraphics[width=4cm]{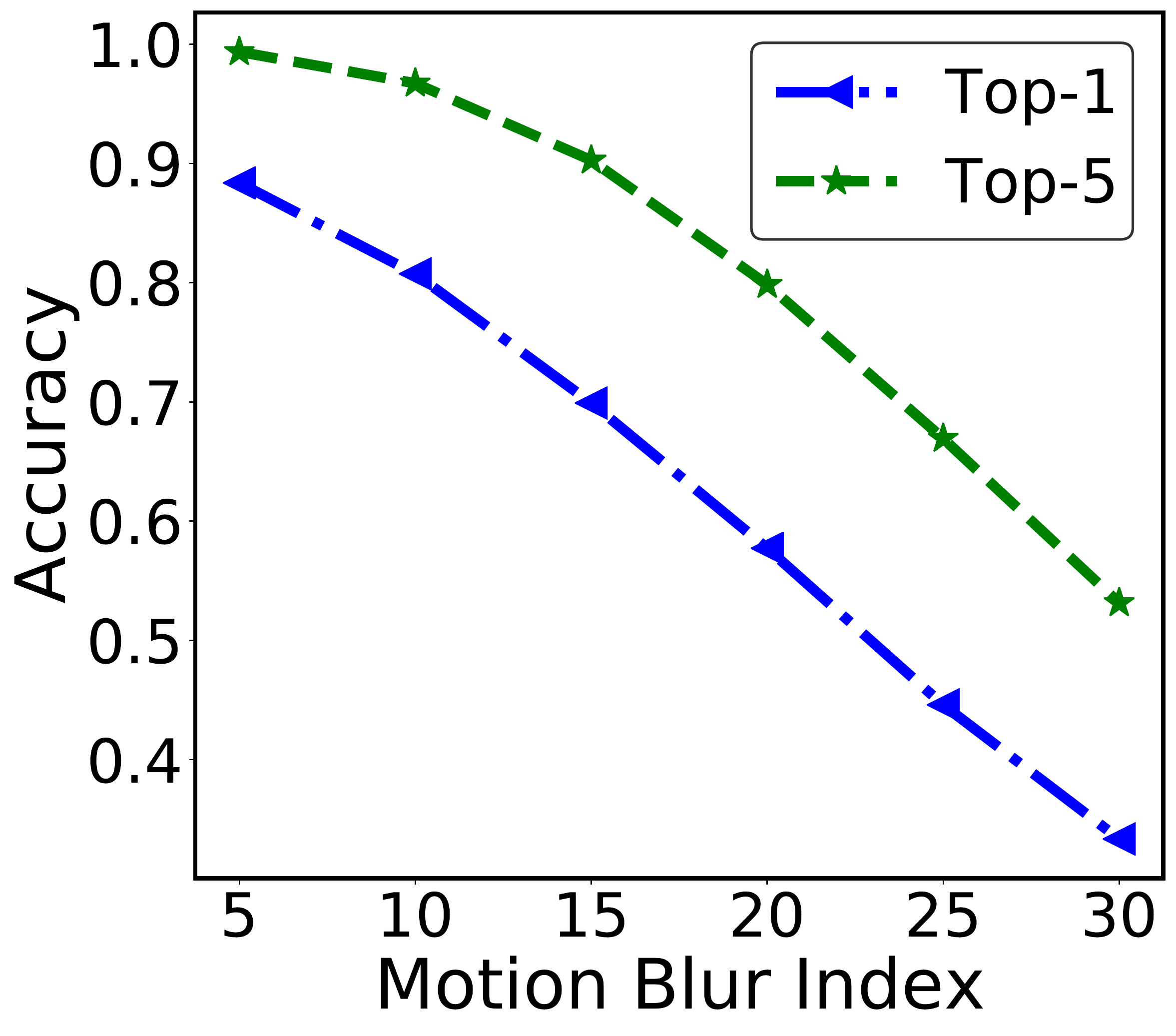}}\\
\subfloat[Compression Artifact]{\raggedleft\includegraphics[width=4cm]{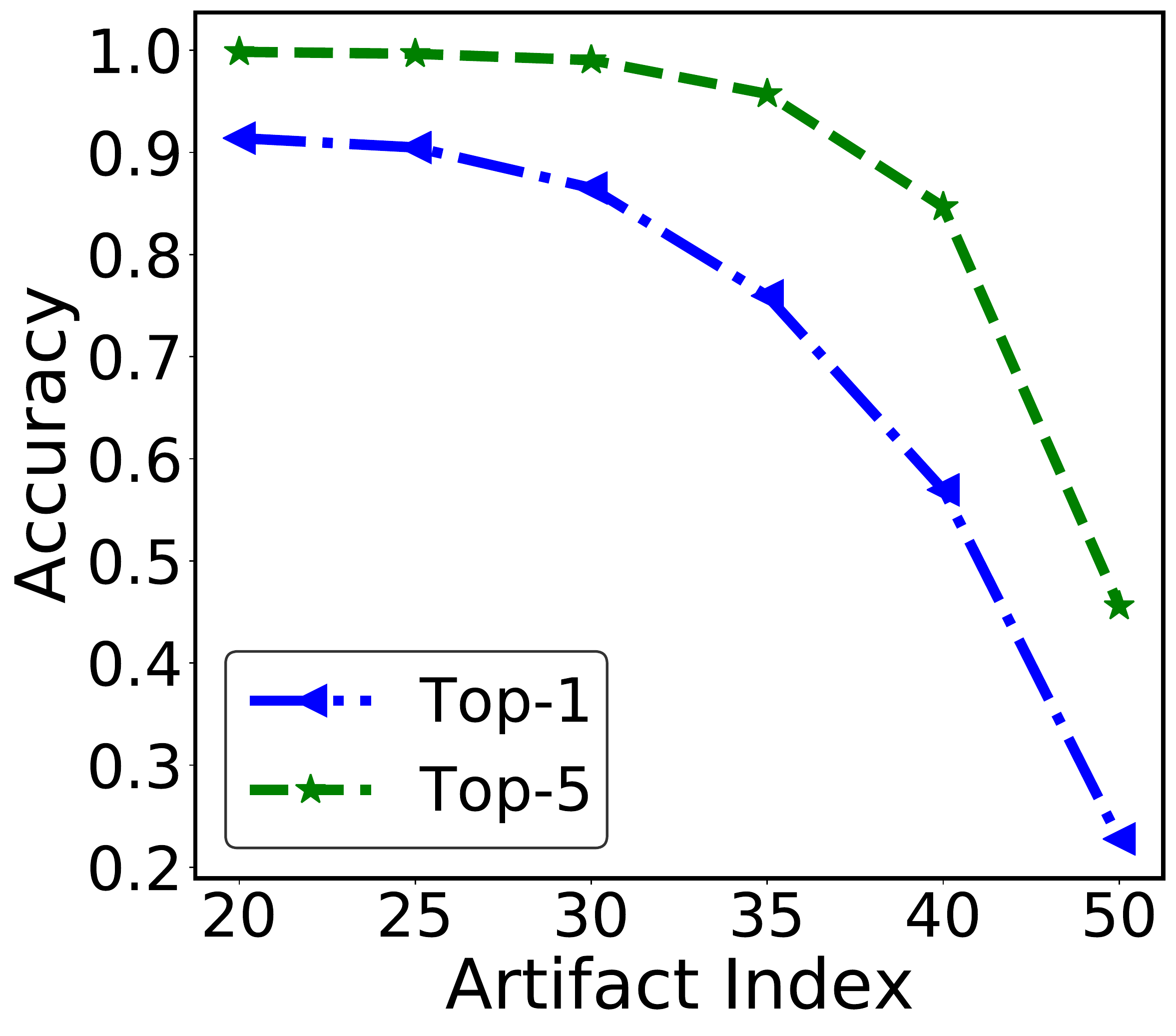}} 
   & \subfloat[Defocus Blur]{\raggedright\includegraphics[width=4cm]{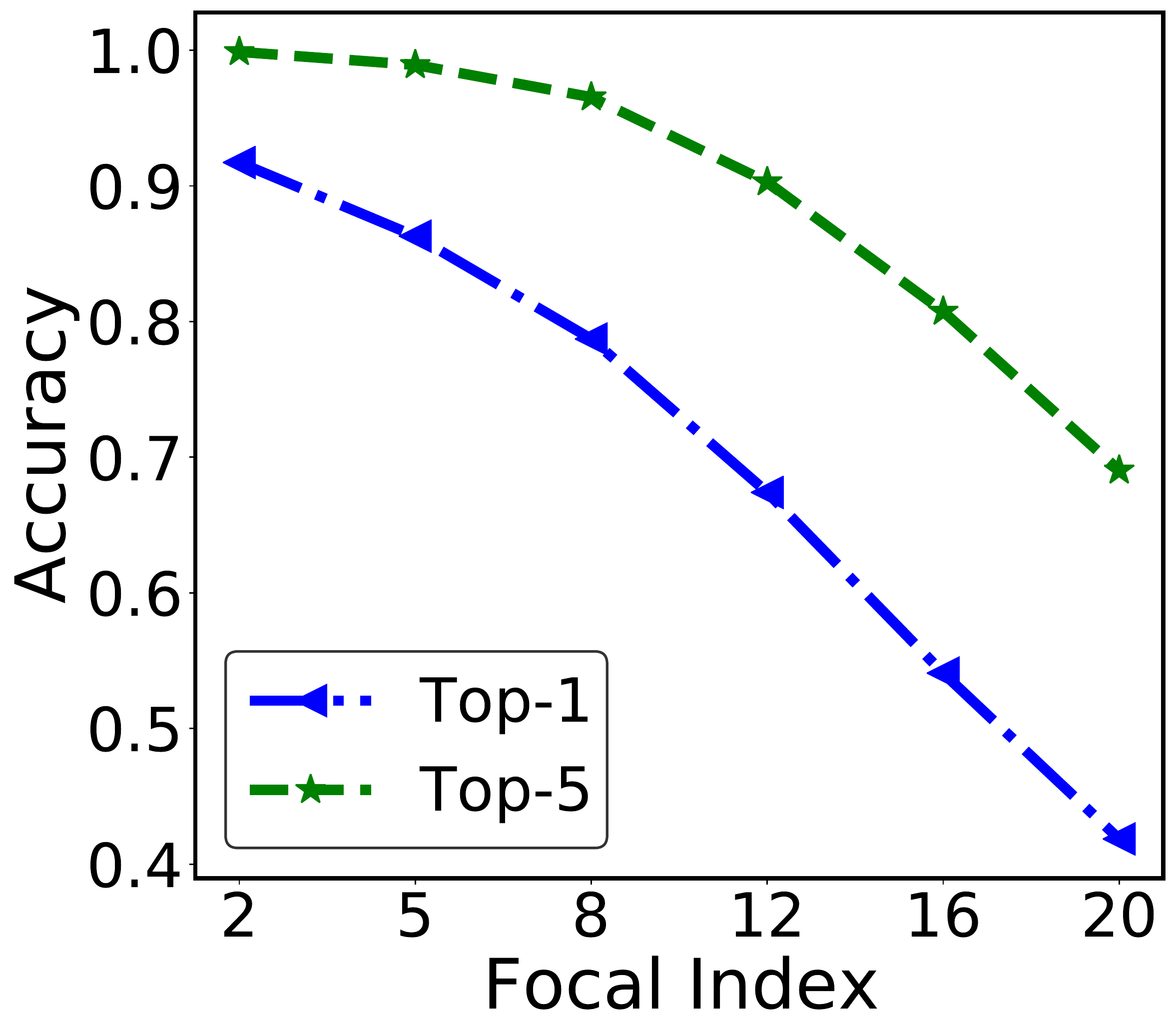}}
\end{tabular}
\end{tabularx}
\caption{Effect of distortions on accuracy}
\label{fig:distortion_effects}
\end{figure}
\secref{section:intro} showed how an image classifier falters, with both a high confidence error and a low confidence error, in the presence of minor distortions. 
To further understand the impact of distortions, we conducted experiments with several image classifiers on a distorted-image dataset, which we created by distorting images from ImageNet. Multiple types of distortions were used, and we also varied the degree of each type of distortion.
We use the top-1 and top-5 accuracy to understand the impact of distortions on classifiers.
\figref{fig:distortion_effects} shows how accuracy is affected by different types and degrees of distortions.

We observe that for any distortion type, as we increase the degree of distortion, the accuracy drops. For example, consider \figref{fig:distortion_effects}(a). As we vary the Brightness index to be either greater than 1 (over exposed) or less than 1 (under-exposed), the classifier's accuracy drops from 90\%, but a precipitous drop is observed for Brightness index beyond 1.25.
In contrast, slight increase in the degree of motion blur or defocus blur leads to an almost linear drop in the classifier's accuracy (\figref{fig:distortion_effects} (b) and (d)). So, the classifier is particularly sensitive to even slight motion or defocus blur, while it can tolerate a modest increase or decrease in brightness related distortion. Our results show that type of distortion, and the degree of distortion, matters, and classifiers can tolerate some type of distortions better than others. This implies that any analytical quality assessor must capture the differential impact of such distortions. Our experiments with four other popular image classifiers, and with more distortion types, also show similar trends.

\subsection{Weak correlation between visual and analytical quality}
\label{subsection:iqa_correlation}
\begin{figure}
    \centering
    \includegraphics[width=\columnwidth]{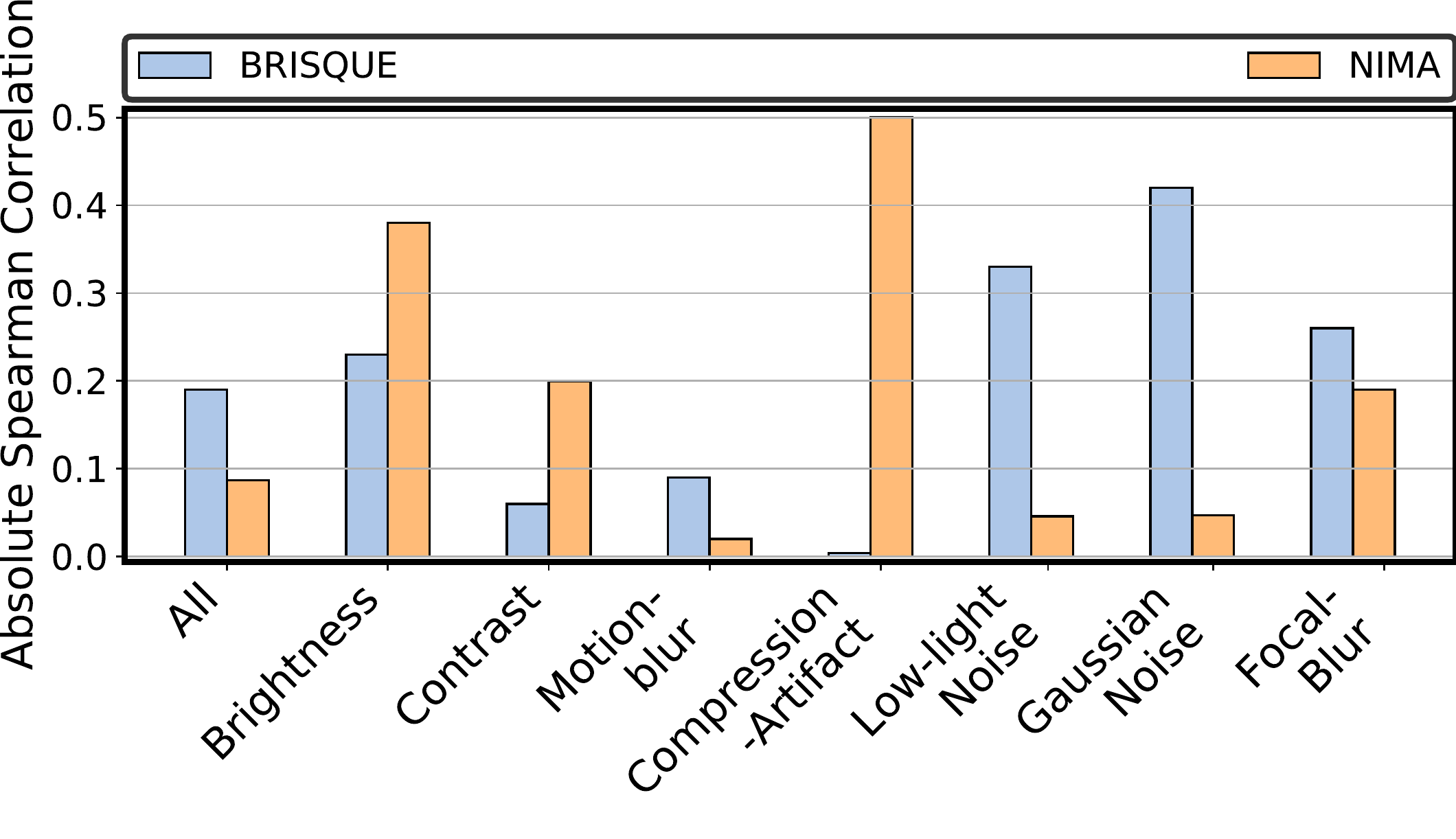}
    \caption{Correlation between state-of-the-art IQA models. ``All" signifies the cumulative result of considering all the different types of distortions.}
    \label{fig:iqa_classifier_correlation}
\end{figure}

In this section, we examine the correlation between perceptual and analytical quality. We use state-of-the-art IQA techniques like NSS-based BRISQUE~\cite{brisque_mittal2012no} and CNN-based NIMA~\cite{talebi2018nima}, which are expressly designed to detect visual quality degradation, to quantify visual quality. We obtain analytical quality scores from a classifier, for a common set of images with various types and degrees of distortion. Here, we define analytical quality of an image as the confidence of correct class (CCC) as observed at the softmax layer of a classifier.

\figref{fig:iqa_classifier_correlation} shows sets of correlation results between perceptual quality scores from each IQA technique, and the analytical quality scores from the classifier, for different types of distortions. For example, consider the two bars for motion blur. For BRISQUE, our experiments show that the absolute Spearman correlation is 0.1, which suggests a weak correlation. Similarly, for NIMA, we see a very weak correlation. We observed a higher correlation between quality scores of NIMA and the classifier (0.5) for compression distortions. Experiments with four other classifiers showed a similar trend. So, we empirically conclude that there is a weak correlation between visual quality and analytical quality, and the extent of correlation depends on the type and degree of distortion. This weak correlation implies that IQA methods are poor estimators of analytical quality of images.

\section{Design}
\label{section:design}

Perceptual IQA methods have 2 essential parts - a feature extractor, which captures important aspects of the image from a perceptual quality point of view, and a regressor, which assigns a quality score. Inspired by the perceptual IQA design, we hypothesize that a good \emph{analytical quality} assessor should have the following desirable properties -
\begin{itemize}
    \item The feature extractor of the assessor should extract features that are representative of the image features that a classifier typically considers. Please note that it does not have to classify images, like a classifier does. So, its feature extraction process does not have to learn higher-level features that are necessary for image classification. 
    \item Any analytical quality assessor must show strong correlation with a classifier's notion of image quality. Therefore, it should consider a classifier's opinion, rather than a perceptual opinion of a human observer. Accordingly, the regressor in the assessor should produce quality scores that correlate well with a classifier's notion of image quality. 
    \item The analytical quality assessor must be efficient, with inference speeds that are much higher (10x or better) than a classifier and a model size that is significantly smaller. This will ensure that the assessor can be used in resource-constrained mobile, embedded and edge processing environments. 
\end{itemize}
We now describe the design of our proposed analytical quality assessor to satisfy the above properties. 

\subsection{Classifier Opinion}
As mentioned in~\secref{subsection:iqa}, training of a perceptual IQA model requires images (original and distorted), and the human opinion scores for each image. The perceptual IQA then learns the mapping between each image and its human opinion.

Our insight is that by replacing the human opinion score with the classifier's opinion score, we can dramatically improve analytical quality assessment. To this end, we discuss two ways of computing a classifier opinion for an image. By using similar ideas, one can devise more elaborate quality scores from the results of a classifier, and the  proposed approach is still applicable to train an effective analytical quality estimator.

\subsubsection{Supervised Classifier Opinion}

Just as human opinion is based on scoring the visual differences between original and distorted images, classifier opinion should also depend on the differences between the original and distorted images. 

\begin{figure}[t]
    \centering
    \includegraphics[width=\columnwidth]{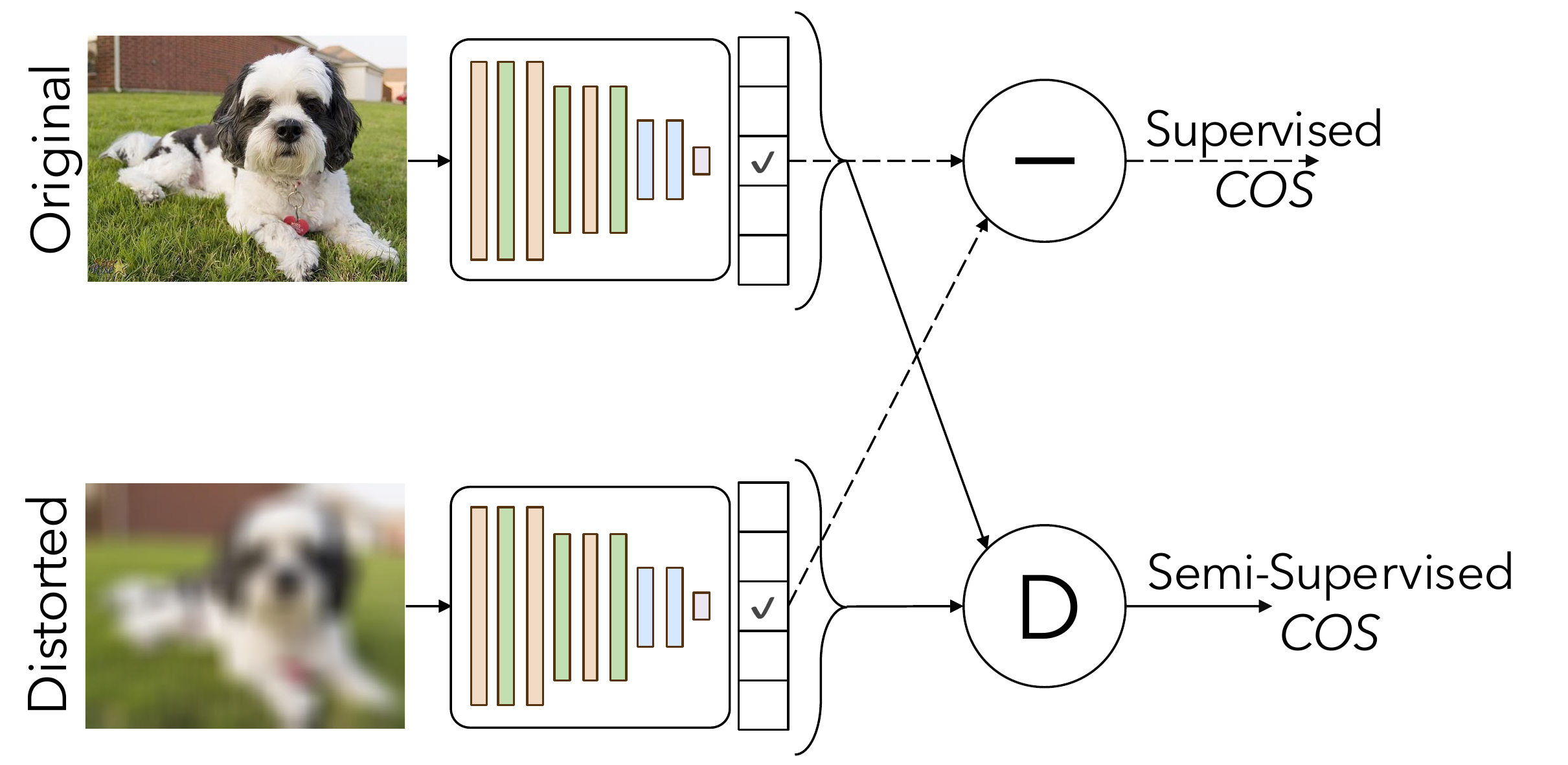}
    \caption{Semi-supervised and supervised Classifier Opinion Score}
    \label{fig:combined_cos}
\end{figure}

We consider the sum of correct class confidence (CCC) and a {\it normalized} correct class rank (NCCR) as an indicator of the analytical quality of an image. The sum takes into account both the correctness, and the confidence in the classification. NCCR maps the correct class rank (CCR) to a real number between 0 and 1, where last rank tends to 0. If the number of classes is $N$, then $NCCR = (N - CCR) / N$. Computation of this sum requires knowledge of the true class of the image, which makes this approach a supervised method. We define the \emph{classifier opinion score} (COS) as the difference between the sums for the original image and its distorted version (as shown in ~\figref{fig:combined_cos}).  Other linear combinations of CCC and NCCR can also be used as the classifier opinion score. 

To attain a more robust opinion, we use several different classifiers, and compute the  mean of the COS scores for the image across the different classifiers. We refer to the mean as MCOS, which is computed using Equation~\ref{eqn:mcos}. 
\begin{empheq}[box=\fbox]{align}
    \scriptstyle MCOS=\frac{1}{\|M\|}\sum_{i\in M}((CCC+NCCR)^i_{org} - (CCC+NCCR)^i_{dist})
\label{eqn:mcos}
\end{empheq}

\subsubsection{Semi-supervised Classifier Opinion}

As mentioned above, computing MCOS requires labeled images. Intuitively, this seems like an excessive requirement, since the goal is not to learn anything specific to a class, but to only capture the classifier's perceived image quality. Therefore, we also propose a semi-supervised approach, which does not require labeled data. We use the entire softmax output of the classifier for a given input image as an indicator of the image quality. Based on the softmax output, we define a new, \textit{semi-supervised} classifier opinion score (COS$_{SS}$), which is the \emph{distance} between the softmax outputs for the original image and its distorted version. The intuition behind this is that softmax output for original images will tend to be a unimodal distribution across the classes, with a strong peak at the correct class. On the other hand, as the image gets distorted, this distribution will either tend towards uniformity as the classifier will not be able to discern classes strongly or tend to be unimodal distribution with a strong peak at a wrong class.

The distance between softmax outputs can be calculated using a number of different methods - KL divergence~\cite{KL}, Mean Absolute Difference (MAD), L1/L2 norms, Bhattacharyya distance~\cite{Bhatt}, JS divergence~\cite{JS}. \figref{fig:combined_cos} shows how semi-supervised COS will be calculated.

Like the supervised case, we use several different classifiers to get a better, more robust classifier opinion. We compute the mean of the COS$_{SS}$ values for the image across different classifiers (see Equation~\ref{eqn:mcos_ss}). We refer to the mean score as MCOS$_{SS}$, which is the classifier opinion score for the semi-supervised case. 
\begin{empheq}[box=\fbox]{align}
    \scriptstyle MCOS_{SS}=\frac{1}{\|M\|}\sum_{i\in M}D(Softmax^i_{org}, Softmax^i_{dist})
\label{eqn:mcos_ss}
\end{empheq}

\subsection{Feature Extraction for Analytical Quality}

Perceptual IQAs use feature extractors like NSS (natural scene statistics) to extract features that a human observer would use to assess perceptual quality.  Similarly, an analytical quality assessor should extract image features that classifiers would use.
We observe that almost all off-the shelf, high-accuracy classifiers share the following:
\begin{enumerate}
\item They use convolution and pooling layers to extract
{\it local} features. Convolution layers use multiple, small
filters within a patch of the image~\cite{gu2018recent,CNN}.

\item The first few layers extract {\it low-level} features like
edges, shapes, or stretched patterns~\cite{gu2018recent,CNN}.
\end{enumerate}


We could use the entire convolution and pooling layers of a classifier for analytical quality, but such a model will have high inference overhead, and it might capture features that are unnecessary for analytical quality. Instead, we propose to use a truncated network, with only layers that capture low-level, object-independent features. This lowers the overhead of feature extraction in our analytical quality estimator, allowing for real-time performance. By using a pre-trained feature extractor, we also dramatically lower the training times.

\subsection{Putting it all together:~\approach}
\begin{figure}[t]
    \centering
    \includegraphics[width=\columnwidth]{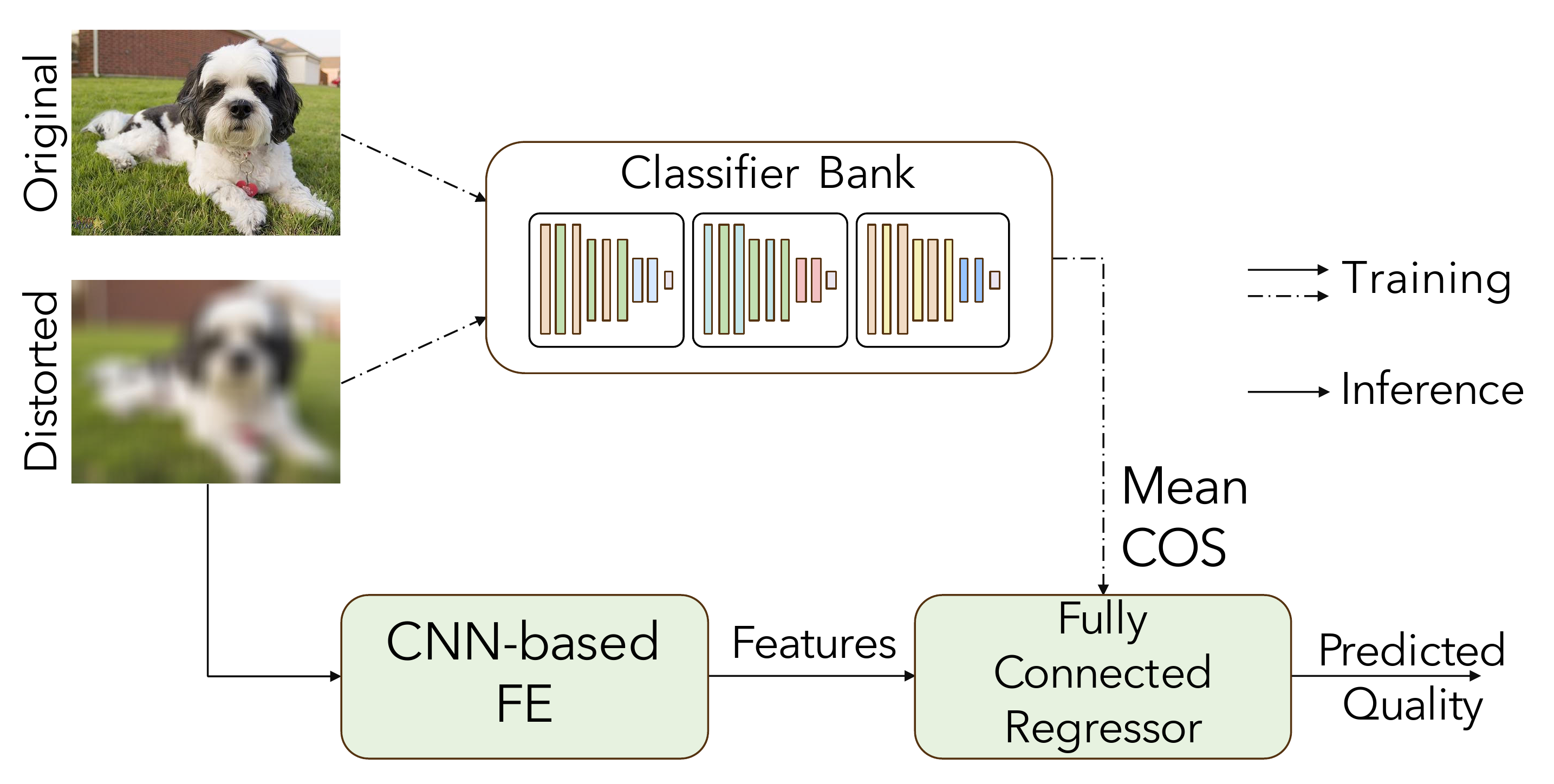}
    \caption{Training and running~\approach.}
    \label{fig:faq_design}
\end{figure}

\figref{fig:faq_design} shows how our analytical quality assessor is built. All the desirable properties described earlier are captured in the different components of the design.
The feature extractor is a shallow, CNN-based pre-trained network. It is lightweight and it extracts exactly the same features that analytical applications use to make decisions.
The regressor is a fully-connected layer that is trained on classifier opinions of images, by using the MCOS or MCOS$_{SS}$ scores.

\section{Evaluation}
\label{section:eval}
Following the recent advancement in edge architectures for AI~\cite{liang2020ai}, edge deployments generally contain GPU or neural accelerators. Here, all performance experiments are conducted on a NVIDIA GeForce RTX 2070 GPU.

\subsection{Training}
\label{subsection:training}
\subsubsection{Training Dataset}
\label{subsubsection:training_dataset}

\begin{table}[h!]
\centering
\begin{tabular}{|c|c|}
\hline
Types of distortion & Range of Distortion \\ \hline\hline
\textit{Brightness} & [0.1,5] \\ \hline
\textit{Contrast} &  [0.1,5] \\ \hline
\textit{Motion-Blur} &  [5,30] \\ \hline
\textit{Compression-Artifact} & [20,50] \\ \hline
\textit{Focal Blur} &  [1,20] \\ \hline
\textit{Gaussian Noise} &  [0.05,0.5] \\ \hline
\textit{Low-light Noise} & [1,100] \\ \hline
\end{tabular}
\caption{Type and range of distortions}
\label{tab:distortion_range}
\end{table}

We train \approach\ by using images from the validation set of ImageNet ILSVRC-2017 dataset~\cite{deng2009imagenet} and their distorted versions. This dataset has 21K original images over 120 different classes. All of these images undergo 7 different types of distortions. For each distortion, 6 degrees of distortion are applied. These degrees are uniformly sampled from within a fixed range. The different distortion types and their ranges are shown in~\tabref{tab:distortion_range}.
80\% of these images from each class are used as input for training, reserving 20\% for testing. Examples of the images are shown in the Appendix.

To obtain MCOS and MCOS$_{SS}$, these training images need to be passed through a bank of classifiers. We use 5 classifiers with least top-1 error on ImageNet validation images~\cite{pytorch_pretrained} - DenseNet-121, ResNeXt-101, Wide ResNet-101, Inception-v3 and VGG-19. 

\subsubsection{\approach\ Model Selection}
\begin{figure}[t]
    \centering
    \includegraphics[width=0.8\columnwidth]{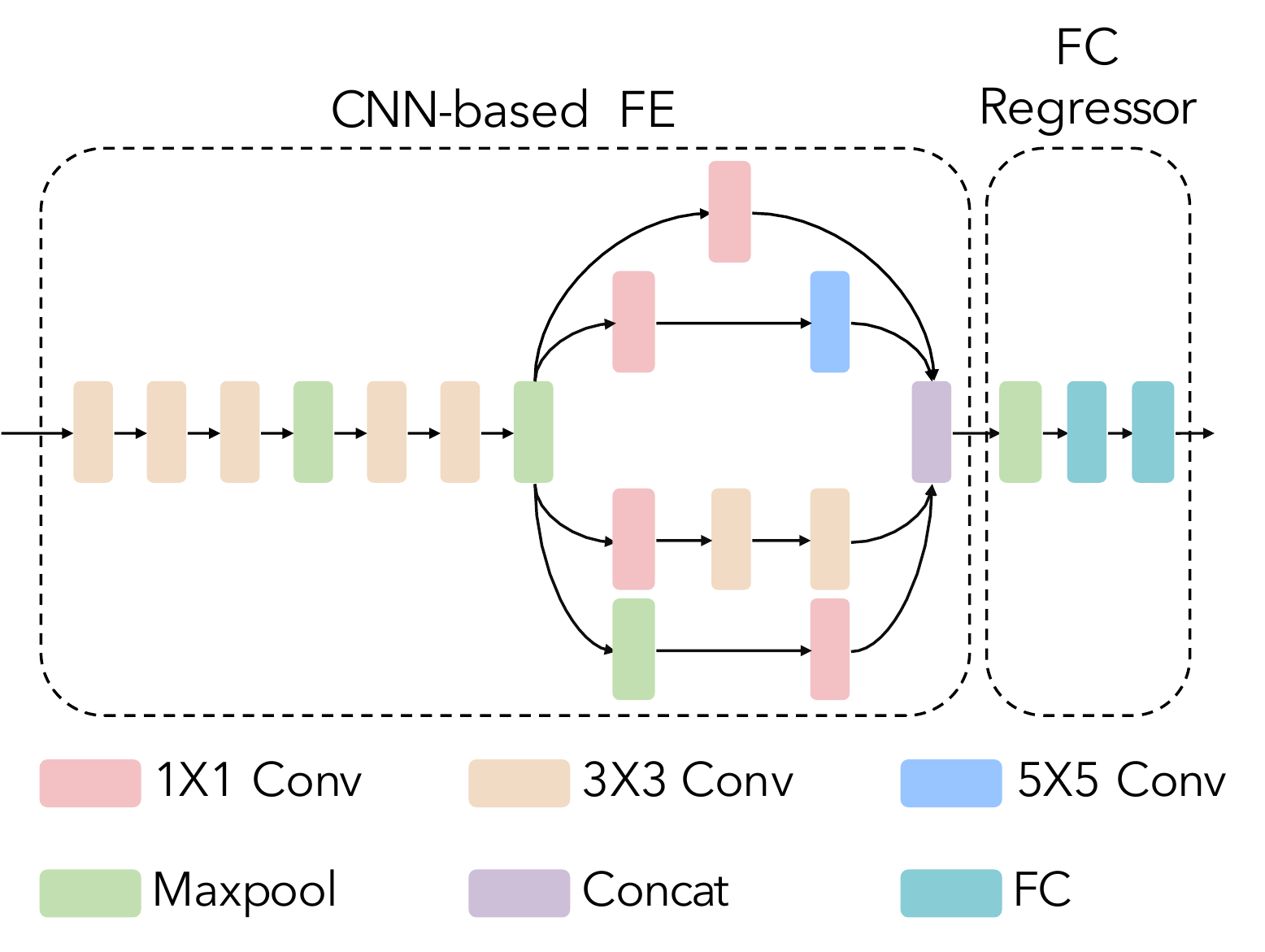}
    \caption{\approach\ model architecture}
    \label{fig:faq_model}
\end{figure}

Different distortions manipulate local statistics at different granularities~\cite{tadros2019assessing}. For example, exposure of light affects coarse textures while motion blur or defocus blur affects finer textures.
In the convolutional layers, larger kernel sizes focus on global textures while stacked convolutional layers extract fine-grained local features. To capture all these granularities, we use the Inception module from \emph{Inception-v3}~\cite{szegedy2016rethinking}, which has convolutional layers with diverse kernel sizes (i.e. 1x1, 3x3 and 5x5) in parallel.
We build the feature extractor for \approach\ using the layers upto the first Inception module in \emph{Inception-v3} followed by a pooling layer. This is followed by fully-connected layers for regression, as shown in~\figref{fig:faq_model}.

\subsubsection{Model Training}
We use transfer learning to train $\approach$.
The Inception based feature extractor is initialized using weights from an ImageNet trained Inception-v3 model, and are frozen. While training, only the weights in the fully connected regression layer are updated. 
The model is trained end-to-end with an initial learning rate of $10^{-5}$ and using the Adam Optimizer~\cite{kingma2014adam} for 200 epochs. Out of 5 different distance measure tested, MAD shows highest monotonic correlation between MCOS$_{SS}$ and distortion levels. Hence, MAD is used to compute the final MCOS$_{SS}$ score.

\subsection{Impact of Design Choices}

We evaluate the impact of using supervised and 
semi-supervised classifier opinion scores, and the proposed lightweight feature extraction model.
The metric for success here is how well the predicted quality score correlates with the classifier's confidence on a large dataset.
We also show the results of using \approach\ as a filter (\approach-Filter) through a Receiver Operating Characteristic (ROC) curve. This shows the discriminating capability of a filter for varying thresholds on quality. Higher area under a ROC curve \emph{(AUC)} indicates better filtering capability.
The definitions of some of the terms are as follows:

\emph{True Positive (TP)}: A frame that is correctly classified by the classifier, and the frame is also passed (i.e., considered to be of good quality) by \approach-Filter. 

\emph{False Positive (FP)}:  A frame that is incorrectly classified by the classifier, but the frame is  passed by \approach-Filter. 

\emph{False Negative (FN)}: A frame that is correctly classified by the classifier but the frame is filtered (i.e., considered to be of poor quality) by \approach-Filter. 

\emph{True Negative (TN)}: A frame that is incorrectly classified by the classifier and the frame is also filtered by \approach-Filter. 

For the experiments in this section, the reserved 20\% images from Section~\ref{subsubsection:training_dataset} are used as the testing dataset. Please note that this dataset contains images that have not been seen by \approach\ during training, and the images have levels of distortions that were not seen in the training dataset. Also, the classifier chosen for assessing confidence-quality correlation was not used in generating any COS scores that were used for training. 
We use ResNet-101~\cite{he2016deep} and GoogLeNet~\cite{szegedy2015going} as the classifiers against which correlation is measured. 
We observed similar results with GoogLeNet.


\begin{figure*}
    \centering
    \begin{subfigure}[t]{0.55\textwidth}
        \vskip 0pt
        \centering
        \includegraphics[width=\textwidth]{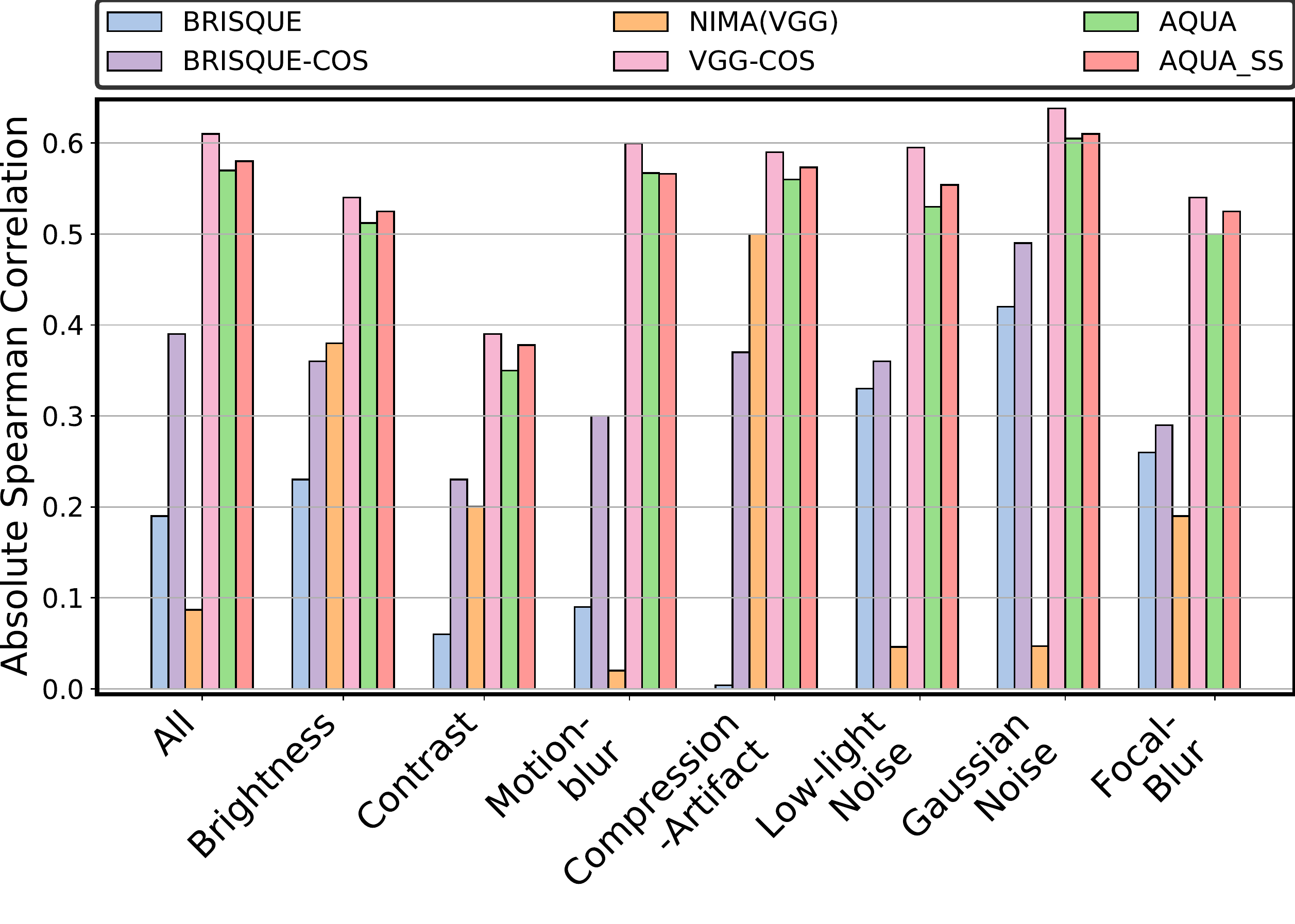}
        \caption{CCC-Quality Correlation}
        \label{fig:impact_correlation}
    \end{subfigure}
    \hfill
    \begin{subfigure}[t]{0.405\textwidth}
        \vskip 0pt
        \centering
        \includegraphics[width=\textwidth]{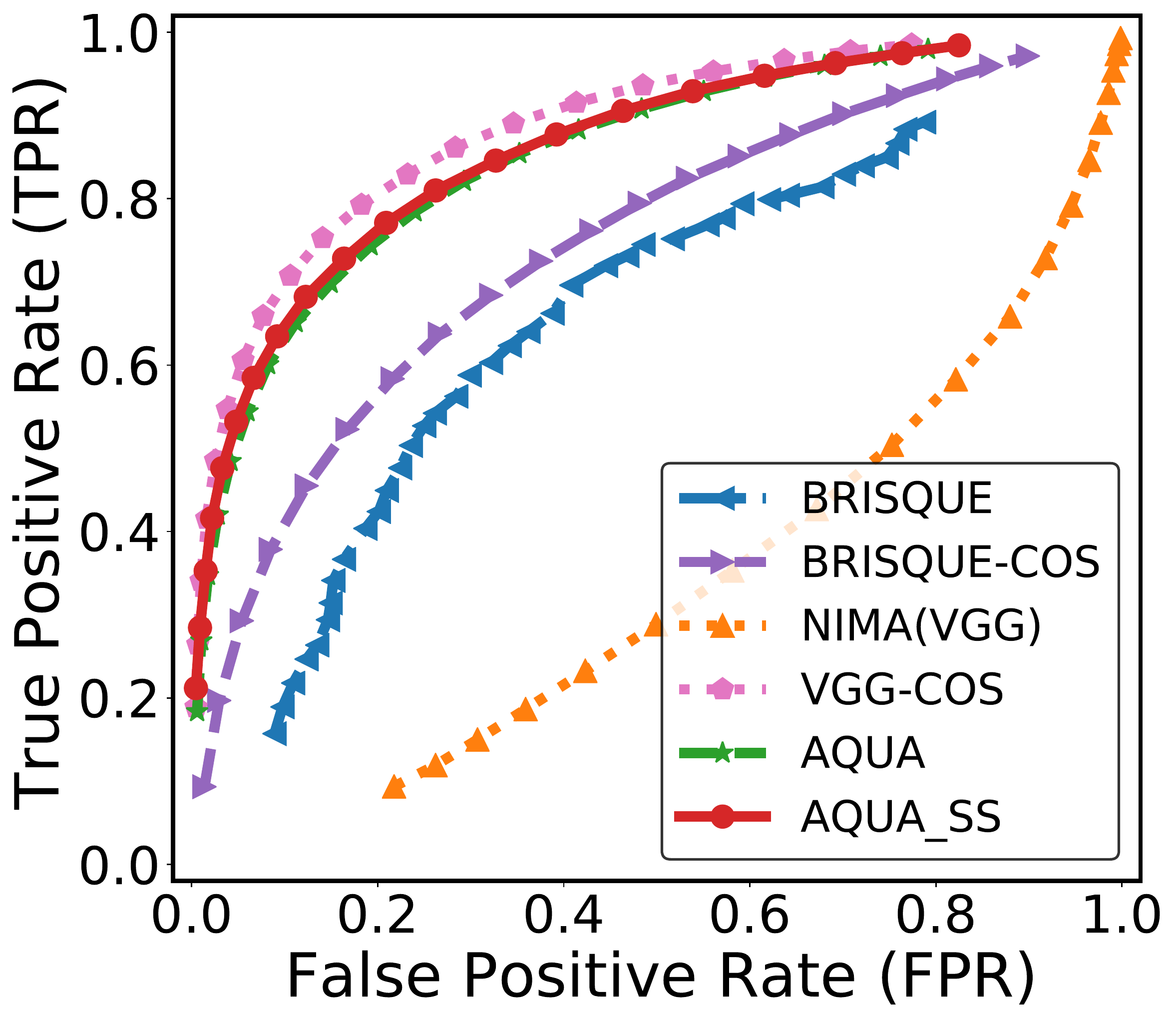}
        \vspace{0.1in}
        \caption{ROC for quality-based filtering}
        \label{fig:roc}
    \end{subfigure}
\caption{Accuracy of different quality assessors}
\vspace{-0.1in}
\end{figure*}

\subsubsection{Impact of Classifier Opinion Score}

To evaluate the impact of the novel training metric, COS, existing visual IQA models were retrained by using COS as the target regression score and the training dataset mentioned in~\secref{subsubsection:training_dataset}. These models were then tested on the reserved 20\% images. The quality scores produced by each model are compared with the test classifier's confidence. The correlation results are shown in \figref{fig:impact_correlation} - the COS versions have the suffix ``COS" appended to their name.\footnote{We use ``VGG-COS" instead of ``NIMA-COS", since NIMA is essentially a VGG-19 model trained for a quality regression task.}

These graphs clearly show that using COS as the training metric drastically improves the correlation of the quality score with the classifier's confidence. This implies that when the COS-trained IQA methods estimate quality of a frame to be low, then it is very likely that the classifier will make a classification error.

The impact of COS is also evident in the ROC curve (\figref{fig:roc}). BRISQUE and NIMA, compared to their COS variants, have lower AUC, and are thus worse filters of analytical quality. 

To understand the impact of semi-supervised training on analytical quality estimates, \approach\ and \approach$_{SS}$ can be compared in both,~\figref{fig:impact_correlation} and ~\figref{fig:roc}. In both of these graphs we see that \approach$_{SS}$ performs similarly or better than \approach. Moreover, since both \approach\ and \approach$_{SS}$ use the same model architecture, the inference speeds are the same as well (Table \ref{tab:system_overhead}).
This shows that semi-supervised training is an effective approach to train an analytical quality assessor. Using the entire softmax output captures more information than just using the correct class, and it also helps the model generalize better.
This will be further established when evaluation on other datasets and applications are presented.

\subsubsection{Impact of Feature Extractor}

The impact of the choice of feature-extractor type and complexity can be seen in \figref{fig:impact_correlation}, which compares the different COS methods. Going from NSS-based feature-extractor in BRISQUE-COS to a CNN-based one in NIMA/VGG-COS provides a significant increase in correlation, almost 2x for some distortions. When we compare the deep feature-extractor in VGG-COS with the shallower extractor in \approach, we see that deep extractors show slightly higher correlation, which is not surprising. The same improvement is also evident in the ROC curves in \figref{fig:roc}. 

The choice of a feature-extractor directly impacts the computation time of the model. \tabref{tab:system_overhead} presents the time for processing a frame. 
Compared to \approach, a deeper feature extractor (VGG) is more than 10x slower and consumes higher GPU memory during computation.
We believe that this disproportionate improvement in latency easily outweighs the minor accuracy advantage of VGG-COS. 

\begin{table}[h!]
\centering
\begin{tabular}{|c|c|}
\hline
Quality Assessor & Latency (ms)     \\ \hline\hline
\textbf{BRISQUE} & 32.3 \\ \hline
\textbf{NIMA} & 175.5 \\ \hline
\textbf{\approach} &  14.2  \\ \hline
\textbf{\approach$_{SS}$} &  14.2 \\ \hline
\textbf{VGG-COS} & 182.2 \\ \hline
\end{tabular}
\caption{Comparing assessor latencies}
\label{tab:system_overhead}
\end{table}


\subsection{Generalizability of \approach}
\begin{figure*}
 \begin{tabularx}{\linewidth}{@{}cXX@{}cXX@{}}
 \begin{tabular}{ccc}
     \subfloat[Object Detection]{\includegraphics[width=5.5cm]{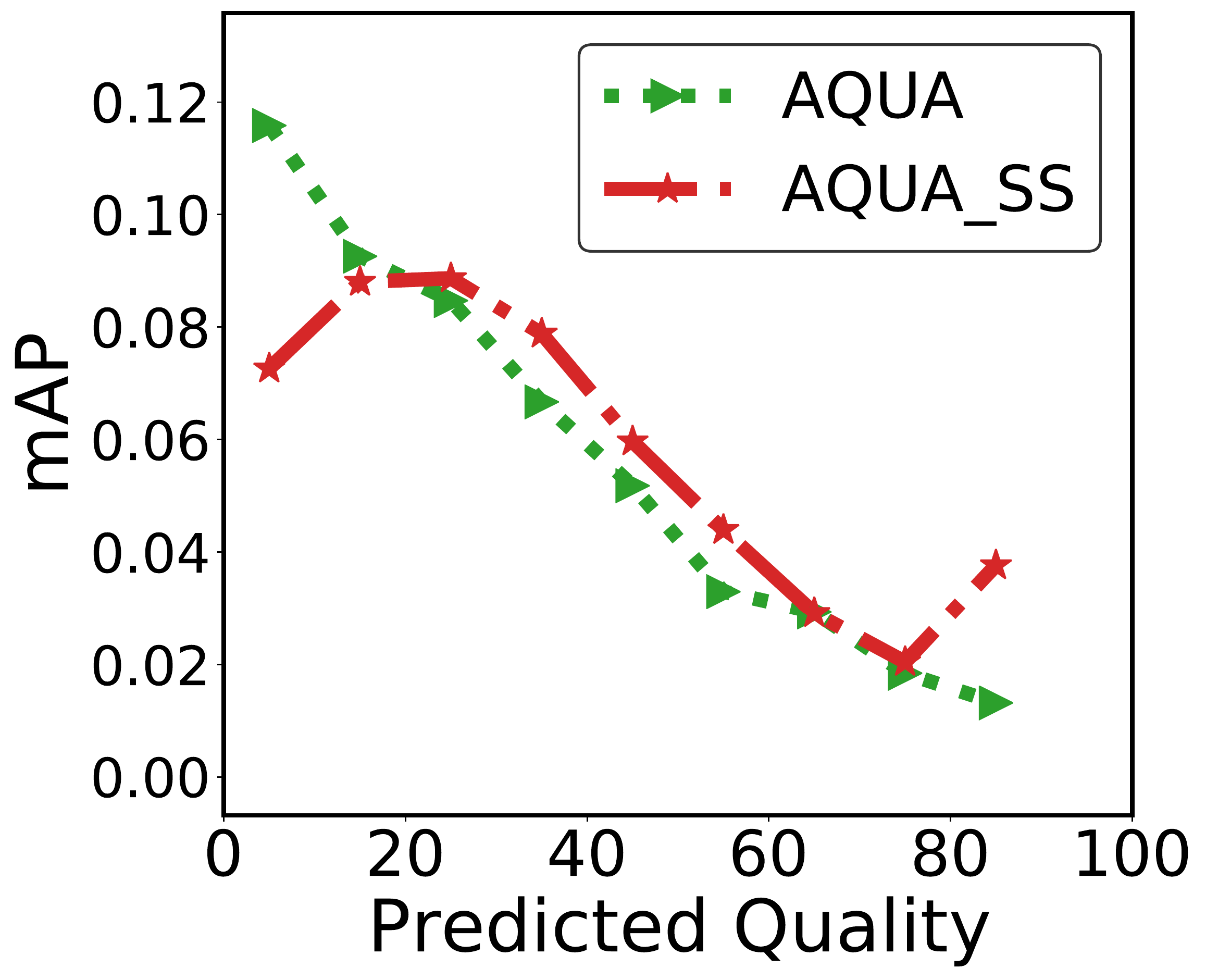}} 
     & \subfloat[Instance Segmentation]{\includegraphics[width=5.5cm]{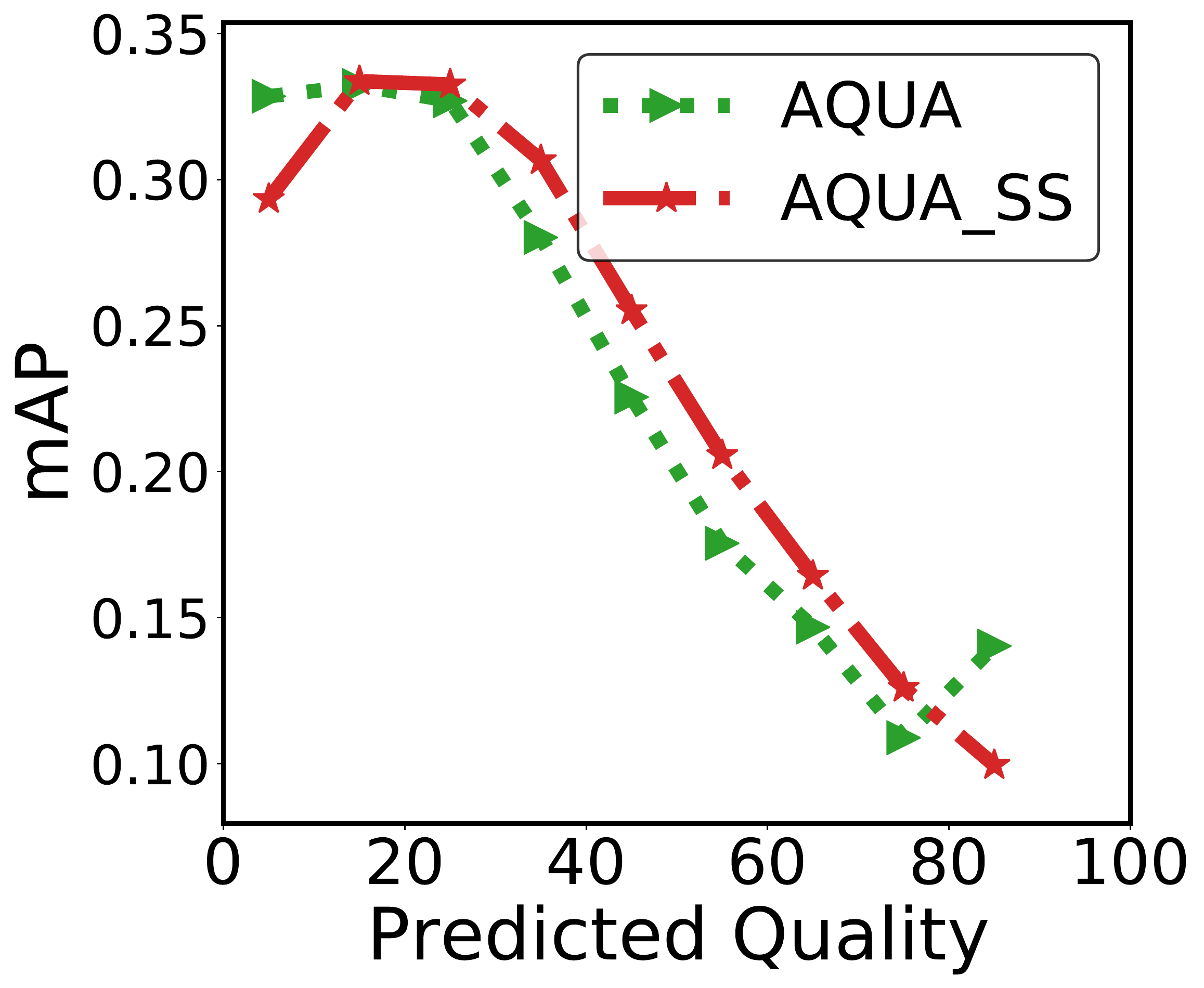}}
     & \subfloat[Keypoint Detection]{\includegraphics[width=5.5cm]{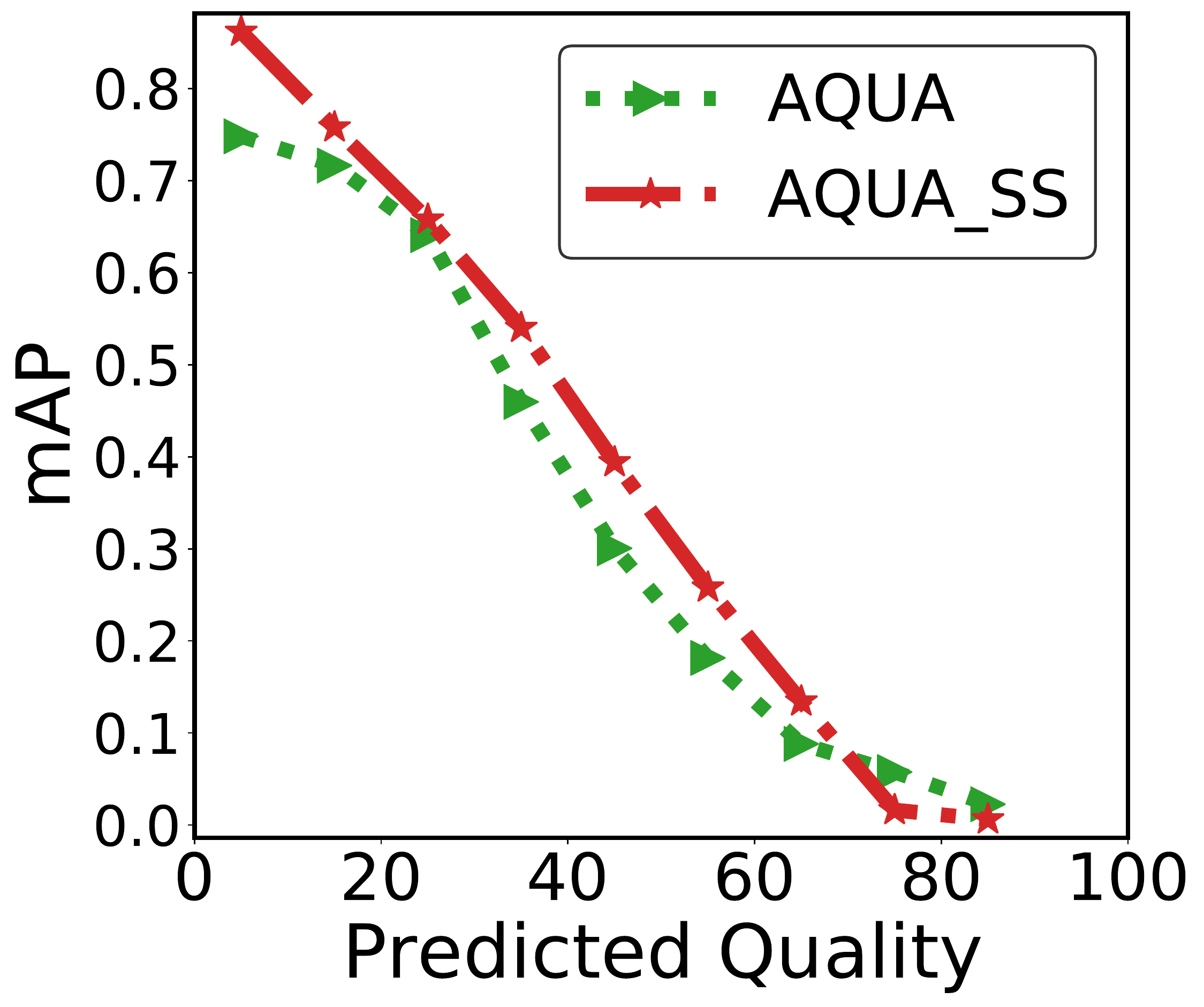}}
 \end{tabular}
 \end{tabularx}
 \caption{\approach\ Generalization}
 \label{fig:faq_generalization}
 \end{figure*}

Edge-based video analytics applications require models for finer-grained tasks, such as object detection for detecting pedestrians or cars, face detection for person recognition, body-keypoint detection for estimating pose and recognizing actions.  
If necessary, AQuA can easily be custom-tailored to each of these different recognizers and detectors. However, in real-world field trials, we observed that many different classifiers and detectors usually process the same video stream as shown in ~\figref{fig:app_mul_analytics}.
Thus a good analytical quality assessor should be able to indicate if any of these models will falter on an input frame.
To establish this, we conducted experiments with models for three different tasks: object detection~\cite{ren2015faster}, instance segmentation~\cite{he2017mask}, and keypoint estimation~\cite{ding2020local}.

The goal of the experiment is to show whether \approach, trained once as described in \secref{subsubsection:training_dataset} on image classifier opinions, can assign quality scores to images that align with other models' accuracy metrics. The metrics to compare are the quality score produced by \approach\ and the mean Average Precision (mAP) \footnote{mAP is defined as mean area under the precision-recall curve for each class. mAP is computed applying pycocotools~\cite{pycoco}} of these models.
The dataset for this experiment is generated by applying distortions, as described in~\secref{subsubsection:training_dataset}, to the COCO dataset~\cite{lin2014microsoft}. 

The results for this experiment can be seen in~\figref{fig:faq_generalization}. 
The graphs show a strong correlation between the mAP and the predicted quality.
That is, if \approach\ estimates that an image is of poor analytical quality, all of the application models under consideration will most likely falter on it. 

Moreover, the generalization of \approach\ is boosted by semi-supervision because it accounts for the frame and the appearance of the object inside the frame, rather than just considering the specific object.

\subsection{Filtering High-confidence Errors}
\label{subsection:filtering_errors}
As discussed earlier, distorted images can cause models to make high-confidence errors. 
We conduct the following experiment to evaluate if an analytical quality-based filter can lower such false-positives (i.e. high-confidence errors). 

We consider the face-recognition application ~\cite{schroff2015facenet}. This application has a database of persons, and each person has one or more images of their face. Given an image, the application either recognizes the faces in the image as known persons (who are already in the database), or it classifies them as unknown. 
\approach\ is placed upstream from the application. If \approach\ thinks the image is of high quality, then the image is forwarded to the application.

 \begin{table}
 \centering
 \begin{tabular}{|c|c|c|}
 \hline
 \textbf{Properties} & FaceScrub & CelebA     \\ \hline\hline
 \textbf{\# of individuals} &  530 & 9211   \\ \hline
 \textbf{\# of test frames } &  18920 & 456100 \\ \hline
 \end{tabular}
 \caption{Face-Recognition Datasets for Evaluation}
 \label{tab:face_recog_dataset}
\end{table}

We use two different face-recognition datasets, CelebA~\cite{liu2015faceattributes} and FaceScrub~\cite{ng2014data} for the evaluation. 
The key properties, i.e.,\ the number of unique faces and total test frames under consideration of these two datasets, are listed in ~\tabref{tab:face_recog_dataset}.

One image per person is used as a reference in the person database. 
For the queries, two images per person are selected from the dataset and are randomly distorted, just as before (\secref{subsubsection:training_dataset}). 

\begin{figure}
 \def\tabularxcolumn#1{m{#1}}
 \begin{tabularx}{\linewidth}{@{}cXX@{}}
 \begin{tabular}{cc}
 \subfloat[FaceScrub]{\includegraphics[width=3.8cm]{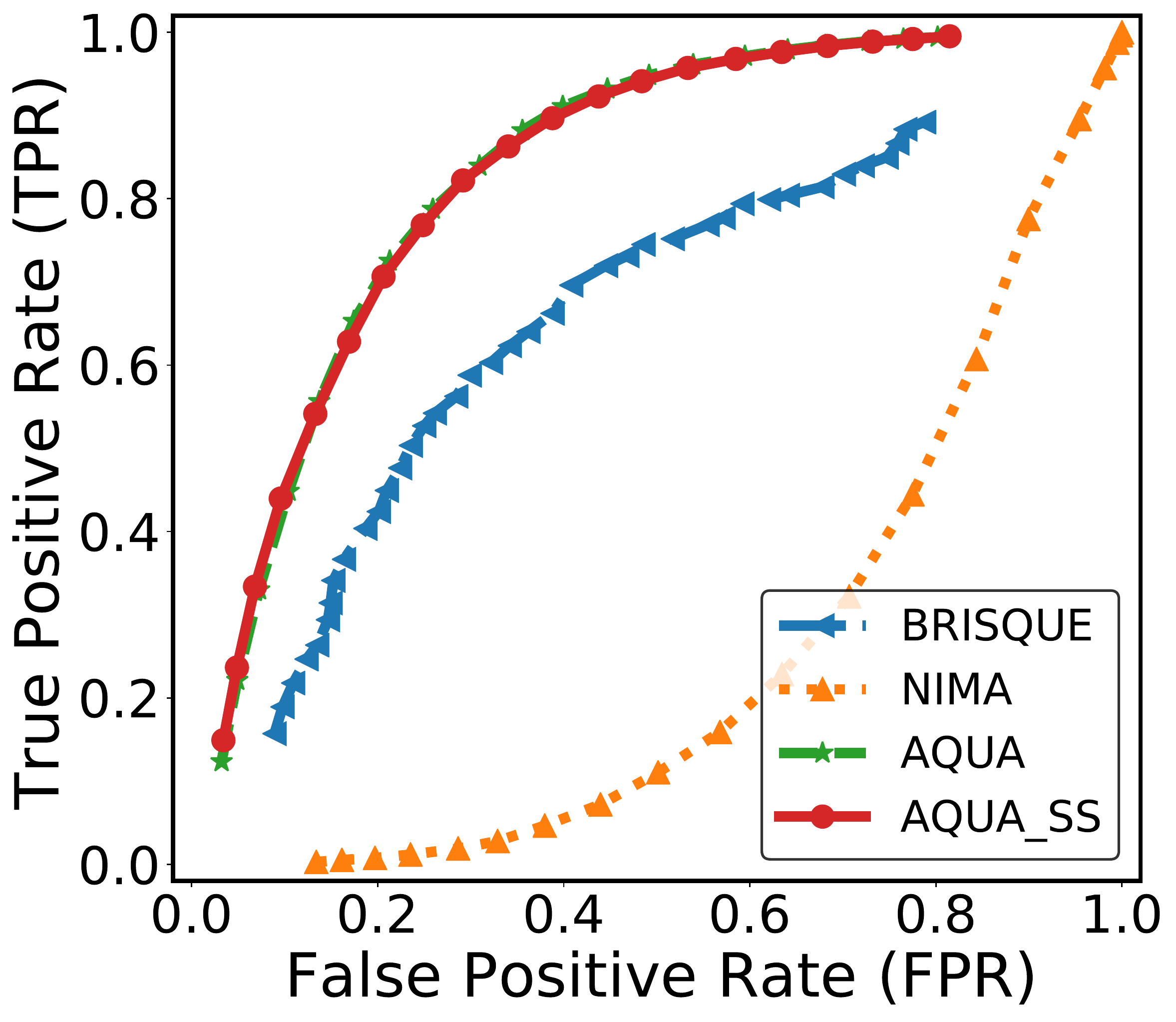}\label{fig:face_rec_roc_facescrub}}
 & \subfloat[CelebA]{\includegraphics[width=3.8cm]{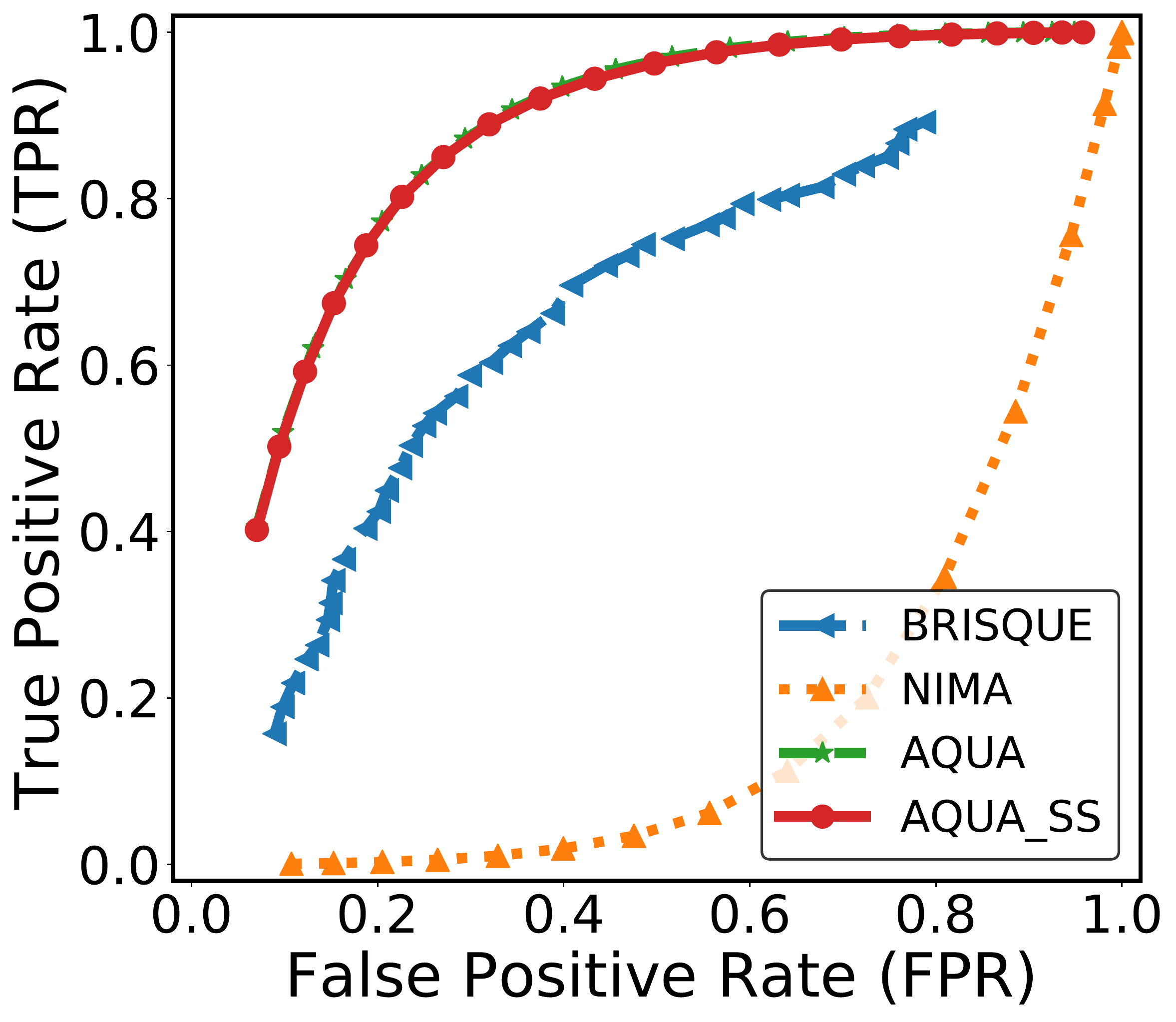}\label{fig:face_rec_roc_celeba}
 }
 \end{tabular}
 \end{tabularx}
 \caption{ROC performance on Face-Recognition Datasets}
 \label{fig:face_rec_roc}
 \end{figure}

\figref{fig:face_rec_roc} shows the results of the experiment on both datasets as ROC curves. The ROC curves show that AUC under either \approach\ or \approach$_{SS}$ is higher than existing IQA methods. This suggests that both \approach\ variants can better filter out poor quality frames as compared to other IQA methods.

\tabref{tab:face_recognition_fp} shows that false-positives (high-confidence errors) reduce by over 17\% when \approach\ is used, with minimal impact on true-positives. 
Thus we conclude that it can effectively lower high-confidence errors in such tasks.

\begin{table}
\centering
\begin{tabular}{|c|c|c|}
\hline
Quality filter & TP Decrease (\%)  & FP Decrease (\%)  \\ \hline\hline
\textbf{BRISQUE} & 0.56 & 6.13  \\ \hline
\textbf{NIMA} & 0.63 & 0.25  \\ \hline
\textbf{\approach} & 0.55 & 17.62 \\ \hline
\textbf{\approach$_{SS}$} &  0.62 & 16.23 \\ \hline
\end{tabular}
\caption{Reduction of FP \& TP for face recognition}
\vspace{-0.3in}
\label{tab:face_recognition_fp}
\end{table}

\subsection{Evaluation on Videos}

\begin{figure}
    \centering
    \begin{subfigure}{\columnwidth}
        \centering
        \includegraphics[width=\textwidth]{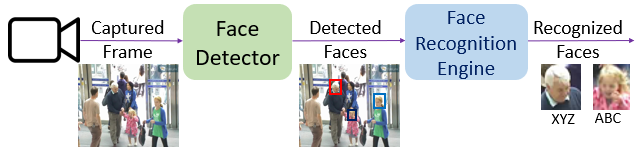}
        \caption{w/o AQuA filter}
        \label{fig:w/oAQuA}
    \end{subfigure}
    \begin{subfigure}{\columnwidth}
        \centering
        \includegraphics[width=\textwidth]{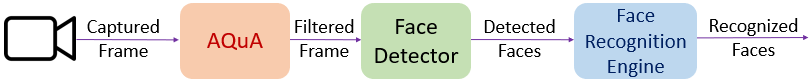}
        \caption{with AQuA filter}
        \label{fig:wAQuA}
    \end{subfigure}
    \caption{Face-recognition pipeline}
\end{figure}

As mentioned earlier, \approach\ can be used as a quality filter, \approach-filter. Such a filter can improve accuracy, and reduce resource usage by filtering out poor quality frames. 
In this section, we evaluate \approach-filter on naturally distorted continuous videos, instead of synthetically distorted images.
Due to the lack of suitable, publicly available video datasets, we used proprietary videos, which have distortions due to environmental changes\footnote{These are confidential customer videos, and hence we cannot share the dataset as of now.}. 
We focus on two videos with multiple chronic distortions, (a) Daytime, and (b) Nighttime.
\begin{itemize}
\item \textbf{Daytime}: This video had the sun shining directly into the camera, blowing out regions of the frame.
\item \textbf{Nighttime}: This video was captured after sunset, and it suffered from low-light noise and under-exposure. 
\end{itemize}

The application running on these video streams is face-recognition, and \approach\ is placed upstream from it, as shown in ~\figref{fig:wAQuA}. The state-of-the-art face-recognition pipeline (as shown in ~\figref{fig:w/oAQuA}) takes captured frames as input from the edge-camera and then pushes them to the face-detector for detecting various face bounding boxes. Each of these face bounding boxes then passes to a face-recognition engine for feature extraction and feature matching with the reference face-database. This face-recognition pipeline is widely used ~\cite{ranjan2019fast,kortli2020face,guo2019survey,sajjad2020raspberry} and also widely adopted by enterprises and governments at airports~\cite{ap1,ap2}, roads~\cite{r1,r2} shopping-mall~\cite{sm1,sm2} for surveillance and enhancing customer experience.


Note, \approach\ has not been trained for this application specifically.
The computation time of the application is: 55ms for face-detection per frame, and 200ms for face feature extraction and matching, per face. \footnote{Our work is currently deployed (field trials) at several major arenas, casinos and airports.}

\begin{figure}
\def\tabularxcolumn#1{m{#1}}
\begin{tabularx}{\linewidth}{@{}cXX@{}}
\begin{tabular}{cc}
\subfloat[Day-time Video]{\includegraphics[width=4cm]{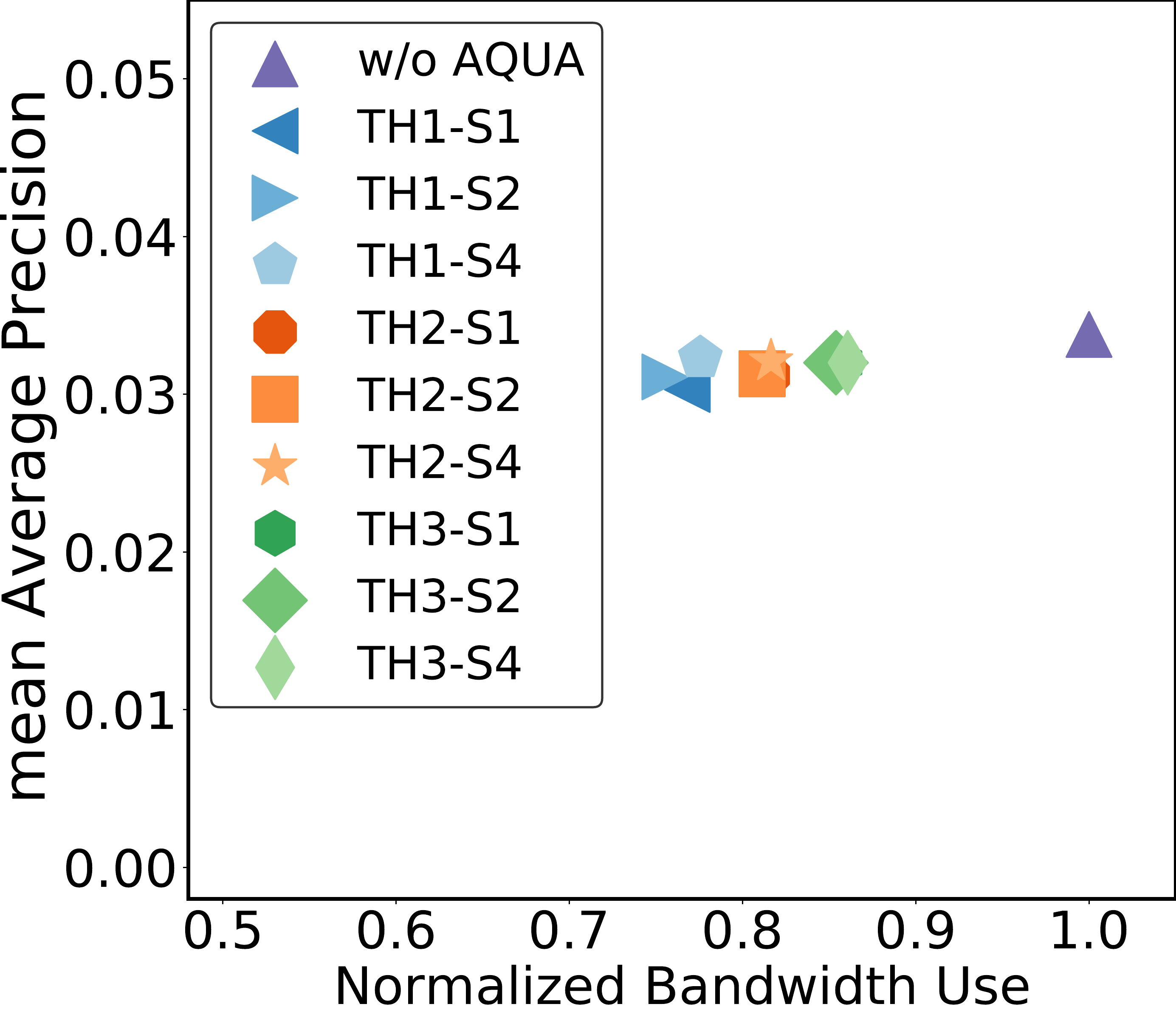}\label{fig:comm_day}} 
   & \subfloat[Night-time Video]{\includegraphics[width=4cm,height=3.45cm]{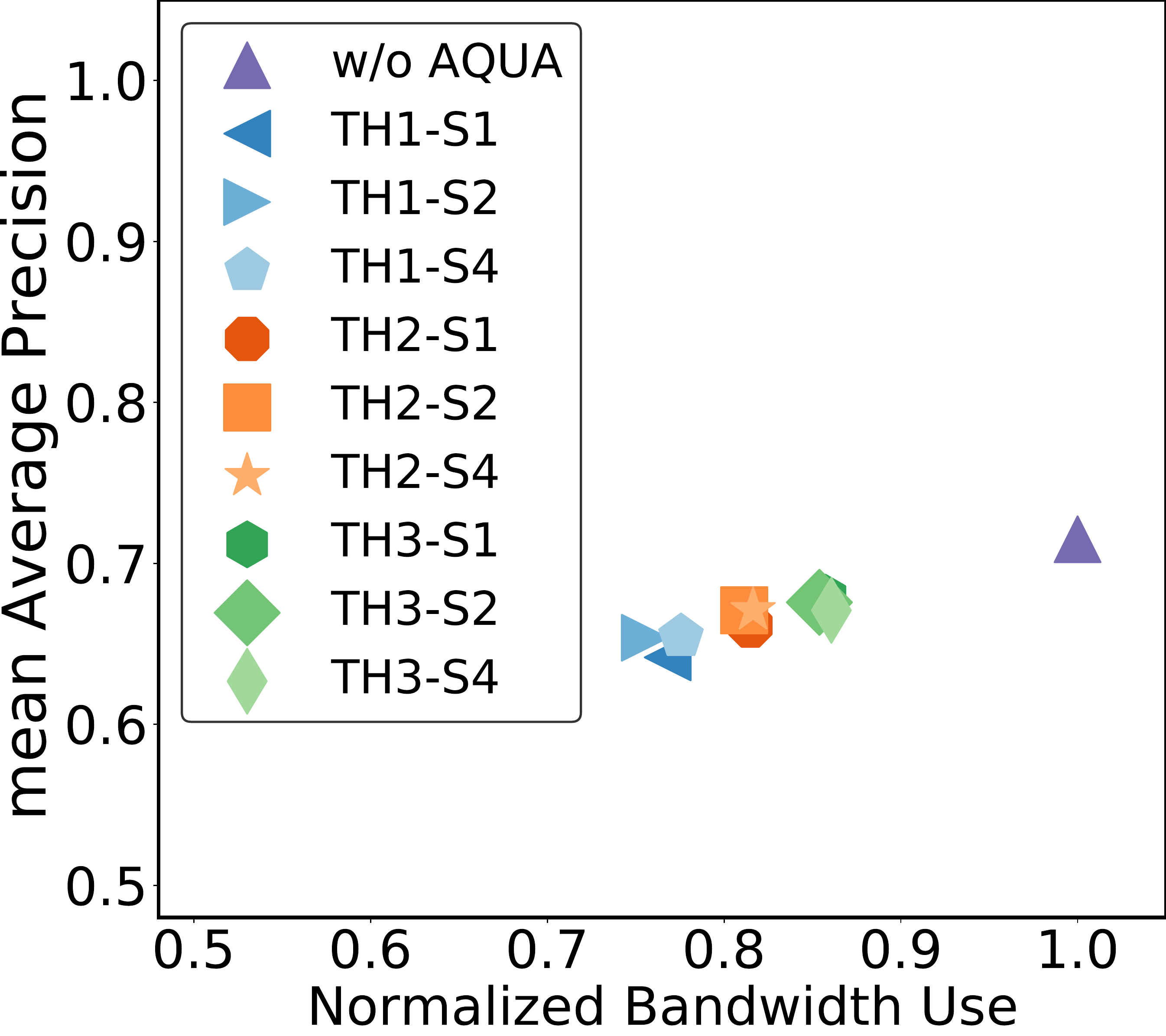}\label{fig:comm_night}}\\
\subfloat[Day-time Video]{\includegraphics[width=4cm]{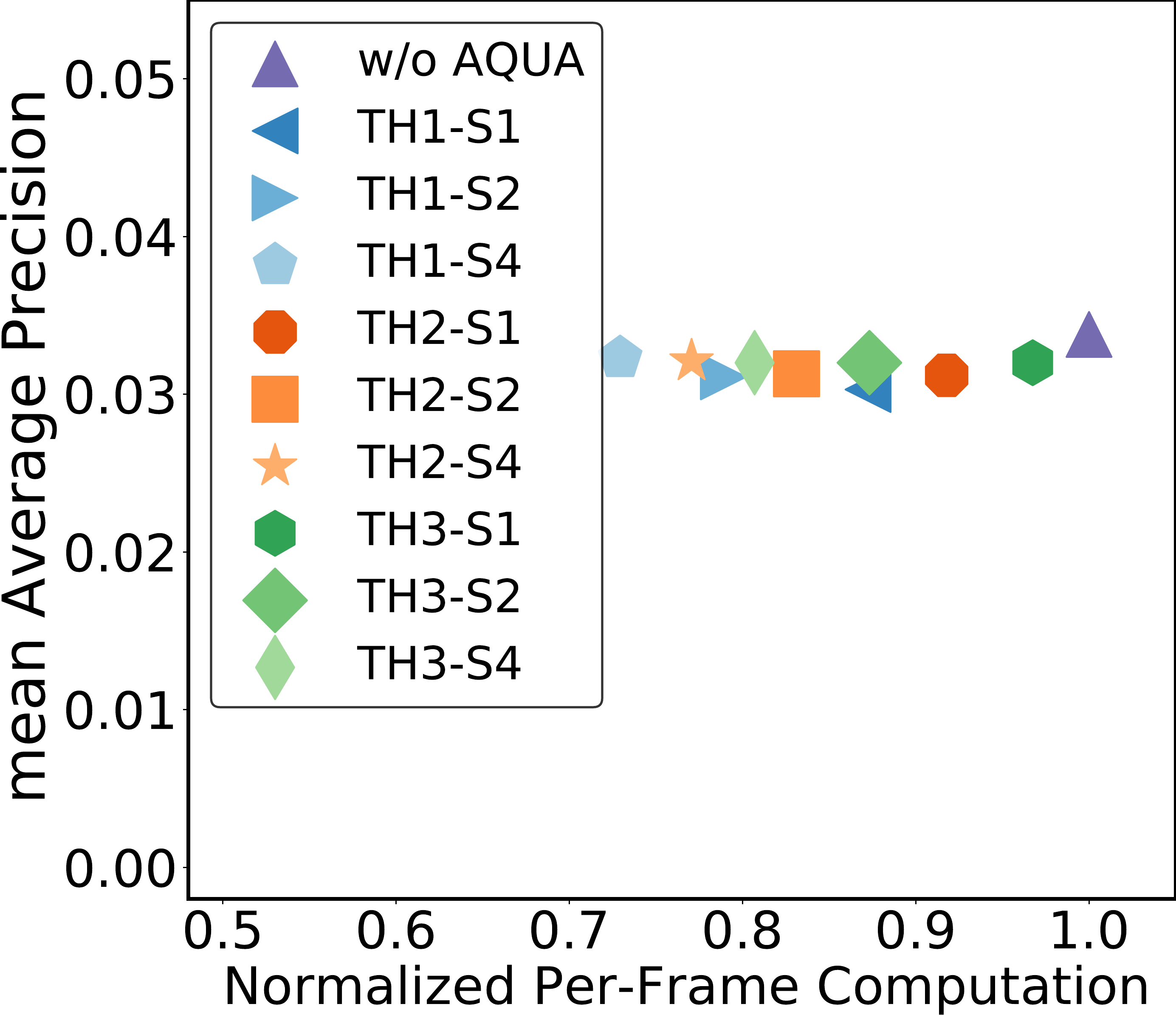}\label{fig:comp_day}} 
   & \subfloat[Night-time Video]{\includegraphics[width=4cm,height=3.45cm]{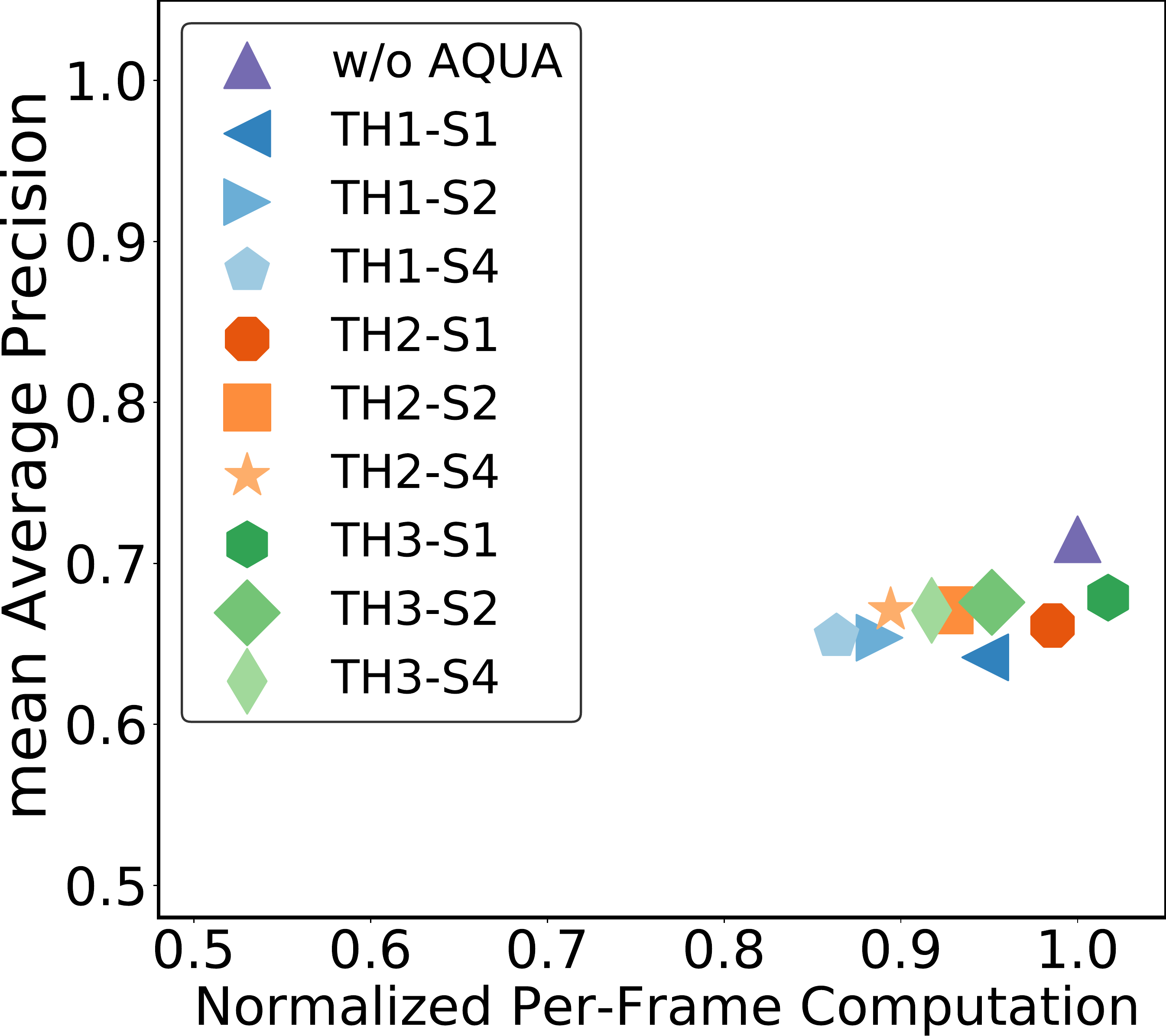}\label{fig:comp_night}}
\end{tabular}
\end{tabularx}
\caption{\approach\ Performance on Videos. TH1, TH2, TH3 signifies three different quality thresholds used for \approach\ while S1, S2, S4 indicates frame sampling for \approach\ prediction}
\vspace{-0.1in}
\label{fig:perf_video}
\end{figure}

To observe the accuracy-resource consumption trade-off, \approach-filter was run with 9 different configurations - three quality thresholds (TH1, TH2, TH3), and three sampling rates (S1, S2, S4). The sampling rate S$x$ denotes that \approach-filter was invoked every $x$ frames. The decision on this frame, whether to filter or not, was then applied to the next $x-1$ frames. 
The metrics collected were mAP, computation time (GPU), and bandwidth consumption. 
The results of the experiments are shown in \figref{fig:perf_video}. Resource consumption has been normalized with respect to the application, i.e. without \approach-filter.


We draw multiple insights from these experiments.

\begin{enumerate}
    \item  Figure~\ref{fig:comm_day}~\ref{fig:comm_night} shows that network bandwidth requirement to stream the captured frames to a remote server reduces by 15\%-28\% under different threshold and sampling settings while minimally reducing the mAP.
    
    \item Figure~\ref{fig:comm_day}~\ref{fig:comm_night} also show that, for the same threshold, sampling doesn't change the bandwidth requirement. This is an artifact of  correlation across adjacent frames. Frame content, quality and hence filtering decision doesn't change abruptly across sequential frames. Even with different sampling rates, the same frames get filtered.

\item However, sampling does help to reduce the computation time, as seen in Figure~\ref{fig:comp_day}~\ref{fig:comp_night}. 
Sampling leverages the temporal locality in videos to help offset the computation overhead of \approach-filter. 

\item If intermediate video feeds are of pristine quality (contrary to daytime video~\footnote{The face-detection accuracy is low for daytime videos due to over-exposure by bright sunlight and glare, which makes face-detection difficult.}, for nighttime video~\figref{fig:comp_night}, there are fewer bad-quality frames from analytical perception), and AQuA quality assessment can be skipped to reduce the computation overhead also shown in \figref{fig:comp_night} through changing sampling rates variable (S1, S2, S4).

\item \textbf{Resource Usage:} The use of \approach-filter reduces the processing resource usage of the face-recognition engine by 27\%. \approach-filter only adds marginal 14ms latency to the original pipeline latency 294 ms (face detector: 55ms, face-recognition:200ms, device-to-edge latency:39ms). This overhead is easily offset by the reduction in processing time of face recognition engine, which now processes fewer frames, and savings in network bandwidth between the device (with \approach) and edge-cloud.

\end{enumerate}
Through these experiments, we also note that \approach-filter can used in the following ways:

\textbf{Edge-only:} For an edge-only scenario~\cite{apicharttrisorn2019frugal}, where all the analytics are performed on the edge device, \approach-filter will reduce the computation overhead by filtering poor quality frames. As more analytics are performed on the same video stream (i.e., multiple face analytics performed on the edge device of Eagleeye~\cite{yi2020eagleeye}) \approach-filter's relative resource reduction will be even higher. A cloud-only system would also benefit in the same way.

\textbf{Edge-Cloud Collaboration:} In edge-assisted real-time AR systems~\cite{yi2020eagleeye, liu2019edge, deng2020fedvision} edge cameras capture the frames and might partially process before streaming to any remote server. Pushing the \approach-filter onto the camera can drop poor quality frames and reduce the streaming bandwidth requirement. This can be applied along with video compression algorithms~\cite{VP9,264} or other filtering approaches~\cite{canel2019scaling,li2020reducto,chen2015glimpse}.

\paragraph{Resource footprint:} \approach-filter only uses layers until the first inception block (only 14\% of the total model size of \textit{Inception-v3}~\cite{szegedy2016rethinking}). Current AQuA model size is 45MB (1.5 GFLOPs), compared to 110MB (256 GFLOPS) for \textit{face-detection RetinaFace model} and 500MB for \textit{face-recognition model}, and AQuA model size can be further reduced by pruning and quantization. However, use of \textit{MobileNetv3}~\cite{howard2019searching} further reduces AQuA model to 10MB without noticeable deterioration of accuracy. These reduced model sizes make \approach\ suitable for edge devices.

\textbf{Scalable Video Analytics:} Along with its filtering capability based on analytical perception, in a multi-camera network systems~\cite{jiang2018chameleon, wang2017scalable, jainspatula}, \approach\ can also enhance the video analytics system’s capability to serve multiple video streams at the same time. For multi-camera video feeds, discarding low-quality distorted frames aids to process multiple parallel streams at the same time. Hence, AQuA can also improve scalability.


\section{Related Work}
\label{section:related}
\subsection{Low Resolution Recognition}
Low resolution (LR) is one of the earliest examples of poor image quality that has been studied in computer vision research.

\subsubsection{Face Recognition}
 One of the first branches of computer vision applications to look into low quality images was face detection and recognition. This is because a number of different security and analytics applications rely on faces, but cameras used for such applications tend to be low resolution, cheap cameras. \cite{DBLP:journals/corr/abs-1805-11519} surveys all recent works in \emph{LR face recognition} and proposes that there still are significant challenges that need to be overcome.

\subsubsection{Super Resolution}
One of the ways to tackle LR images is to construct a high resolution version of them through super resolution. 
This is a classic computer vision problem, which has gotten renewed attention due to the success of deep learning and CNN based models. 

\cite{yang2014single} provides a good overview of different approaches for generic super resolution. There has been additional work to direct super resolution for specific applications, such as, for person identification in a crowded scene, Eagleeye~\cite{yi2020eagleeye}, for object detection~\cite{DBLP:journals/corr/abs-1803-11316}, 
Most recent edge-assisted face-recognition system,  also employs super-resolution to identify missing person accurately from captured LR faces in a crowded urban space.

Although low resolution can have impact on quality of images, it is orthogonal to the kind of distortions considered in this work. Moreover, the methods to overcome it are complementary to this work.

\subsection{Classification on Distorted Images}
Image classification models has recently surpassed human-level accuracy on large datasets such ImageNet~\cite{deng2009imagenet}. This has been made possible through deep learning and CNNs~\cite{krizhevsky2012imagenet,he2016deep,szegedy2016rethinking,szegedy2015going,vgg19simonyan2014very}. However, it has been shown that these models are brittle and can lead to erroneous predictions even when the input is distorted in minimal ways. 

\subsubsection{Adversarial Distortion}
Adversarial distortions are small, calculated and deliberate perturbations on the input images, which are visually imperceptible, that cause classifiers to fail~\cite{goodfellow2014explaining, carlini2017adversarial, athalye2018synthesizing}. 

Although this work addresses distortions too, it looks at addressing ``natural" distortions due to image acquisition or transmission.
\vspace{-0.11in}
\subsubsection{Non-adversarial Distortion}
There have been multiple efforts that show that image classification suffers on images that undergo common distortions, like blur, noise, over-exposure~\cite{pei2018effects,roy2018effects,tadros2019assessing,dodge2016understanding}. The main reason, proposed in these papers, is that most classifiers are trained on high-quality images, typically scraped from the Internet, and hence fail on low-quality images.
\cite{DBLP:journals/corr/VasiljevicCS16} show that fine-tuning an existing classifier with blurred images can improve the classifier's performance, but can impact it's performance on pristine images. \cite{zhou2017classification} showed that classifiers can also be fine-tuned on noisy images using the same approach, but had the same drawback of reduced overall accuracy.

\cite{DBLP:journals/corr/DiamondSBWH17} attaches another network to the input of a classifier that rectifies blurry and noisy images, and thus presents a cleaner image to the classifier. This network, however, requires camera parameters to be trained, making it difficult to generalize. 
\cite{dodge2018quality} introduces MixQualNets, which takes an ensemble learning approach. Each model within the ensemble is an image classifier, but trained with different kinds of distortions. Specifically, their proposed ensemble consists of 3 image classifiers: clean images, noisy images and blurry images. The overall accuracy on all kinds of images is better than each individual classifier, though it comes at a high computation cost.

\cite{zheng2016improving} proposes a new training method, \emph{stability training}, which improves the resilience of the network to common distortions. They validate the approach on highly compressed JPEG images and show that their method outperforms the base image classifier.

Although multiple approaches have been explored to tackle the issue of degraded images, most of these cannot be applied in large scale video-analytics deployments. Most the earlier work requires retraining of classifiers, and since a single application can have a number of different  models, this might not be practical. Most approaches also increase the size of the network, thus leading to higher compute times, which is detrimental to video-analytics applications.

\approach\ takes a different approach, in which it filters degraded frames, safeguarding the accuracy of all models in the application pipeline. Moreover, by filtering such frames at the head of pipeline, it reduces resource usage.

\subsection{Configurable Analytics Systems}
Recently, tuning the video analytics pipeline for better accuracy along with efficient resource usage has gained a lot of attention. Most of these works, e.g.,\ Chameleon~\cite{jiang2018chameleon}, AWStream~\cite{zhang2018awstream}, and
VideoStorm~\cite{zhang2017live}, focus on parameter tuning of \emph{resolution, frame-rate and analytical model under consideration} to achieve better resource-accuracy trade-off. Such parameter tuning are applied after frame acquisition/registration by edge camera. However, during frame acquisition through camera, if inferior-quality frames are continuously captured due to camera misconfigurations (i.e., camera focus, exposure settings), high resource consumption will happen without any desired anlaytics being performed. 
$\approach$  not only reduces redundant resource consumption on edge devices through segregating inferior-quality frames, it can also be trained to predict the misconfigurations that caused quality-deterioration. This will further reduce the chances of inferior-quality frame acquisition.
\section{Conclusion}
\label{section:conclusion}
Live, real-time video analytics applications at edge are increasing, propelled by deep learning and 5G connectivity.
 However, in-camera and transmission distortions cause applications to falter and produce erroneous analytical outcomes.
In this work, we introduced \approach\ to judge frames and assign an analytical quality score. We also proposed that \approach\ can be used as a filter to drop low quality frames with distortion, and eliminates frames that can lead to high-confidence errors. Our approach is inspired from IQA methods, but instead human opinions, we define a new metric, classifier opinion score, that helps train \approach. \approach\ uses a truncated Inception-v3 for feature extraction, to extract low-level, object-independent features.
We evaluate \approach\ and show that it can outperform SOTA IQA methods in terms of correlating with application confidence, filtering and reducing false-positives. We also show that \approach\ generalizes to multiple different video analytics applications and can reduce resource (i.e. communication and compute) consumption as well without degrading inference accuracy while processing video data at scale.

\bibliography{references}


\begin{thebibliography}{84}


\ifx \showCODEN    \undefined \def \showCODEN     #1{\unskip}     \fi
\ifx \showDOI      \undefined \def \showDOI       #1{#1}\fi
\ifx \showISBNx    \undefined \def \showISBNx     #1{\unskip}     \fi
\ifx \showISBNxiii \undefined \def \showISBNxiii  #1{\unskip}     \fi
\ifx \showISSN     \undefined \def \showISSN      #1{\unskip}     \fi
\ifx \showLCCN     \undefined \def \showLCCN      #1{\unskip}     \fi
\ifx \shownote     \undefined \def \shownote      #1{#1}          \fi
\ifx \showarticletitle \undefined \def \showarticletitle #1{#1}   \fi
\ifx \showURL      \undefined \def \showURL       {\relax}        \fi
\providecommand\bibfield[2]{#2}
\providecommand\bibinfo[2]{#2}
\providecommand\natexlab[1]{#1}
\providecommand\showeprint[2][]{arXiv:#2}

\bibitem[\protect\citeauthoryear{??}{DBL}{2018}]%
        {DBLP:journals/corr/abs-1805-11519}
 \bibinfo{year}{2018}\natexlab{}.
\newblock \showarticletitle{Face Recognition in Low Quality Images: {A}
  Survey}.
\newblock \bibinfo{journal}{\emph{CoRR}}  \bibinfo{volume}{abs/1805.11519}
  (\bibinfo{year}{2018}).
\newblock
\showeprint[arxiv]{1805.11519}
\urldef\tempurl%
\url{http://arxiv.org/abs/1805.11519}
\showURL{%
\tempurl}


\bibitem[\protect\citeauthoryear{Apicharttrisorn, Ran, Chen, Krishnamurthy, and
  Roy-Chowdhury}{Apicharttrisorn et~al\mbox{.}}{2019}]%
        {apicharttrisorn2019frugal}
\bibfield{author}{\bibinfo{person}{Kittipat Apicharttrisorn},
  \bibinfo{person}{Xukan Ran}, \bibinfo{person}{Jiasi Chen},
  \bibinfo{person}{Srikanth~V Krishnamurthy}, {and} \bibinfo{person}{Amit~K
  Roy-Chowdhury}.} \bibinfo{year}{2019}\natexlab{}.
\newblock \showarticletitle{Frugal following: Power thrifty object detection
  and tracking for mobile augmented reality}. In
  \bibinfo{booktitle}{\emph{Proceedings of the 17th Conference on Embedded
  Networked Sensor Systems}}. \bibinfo{pages}{96--109}.
\newblock


\bibitem[\protect\citeauthoryear{Athalye, Engstrom, Ilyas, and Kwok}{Athalye
  et~al\mbox{.}}{2018}]%
        {athalye2018synthesizing}
\bibfield{author}{\bibinfo{person}{Anish Athalye}, \bibinfo{person}{Logan
  Engstrom}, \bibinfo{person}{Andrew Ilyas}, {and} \bibinfo{person}{Kevin
  Kwok}.} \bibinfo{year}{2018}\natexlab{}.
\newblock \showarticletitle{Synthesizing robust adversarial examples}. In
  \bibinfo{booktitle}{\emph{International conference on machine learning}}.
  PMLR, \bibinfo{pages}{284--293}.
\newblock


\bibitem[\protect\citeauthoryear{Bhattacharyya-distance}{Bhattacharyya-distance}{[n.d.]}]%
        {Bhatt}
\bibfield{author}{\bibinfo{person}{Bhattacharyya-distance}.}
  \bibinfo{year}{}\natexlab{}.
\newblock
  \bibinfo{howpublished}{\url{https://en.wikipedia.org/wiki/Bhattacharyya_distance}}.
\newblock


\bibitem[\protect\citeauthoryear{Canel, Kim, Zhou, Li, Lim, Andersen, Kaminsky,
  and Dulloor}{Canel et~al\mbox{.}}{2019}]%
        {canel2019scaling}
\bibfield{author}{\bibinfo{person}{Christopher Canel}, \bibinfo{person}{Thomas
  Kim}, \bibinfo{person}{Giulio Zhou}, \bibinfo{person}{Conglong Li},
  \bibinfo{person}{Hyeontaek Lim}, \bibinfo{person}{David~G Andersen},
  \bibinfo{person}{Michael Kaminsky}, {and} \bibinfo{person}{Subramanya~R
  Dulloor}.} \bibinfo{year}{2019}\natexlab{}.
\newblock \showarticletitle{Scaling video analytics on constrained edge nodes}.
\newblock \bibinfo{journal}{\emph{arXiv preprint arXiv:1905.13536}}
  (\bibinfo{year}{2019}).
\newblock


\bibitem[\protect\citeauthoryear{Carlini and Wagner}{Carlini and
  Wagner}{2017}]%
        {carlini2017adversarial}
\bibfield{author}{\bibinfo{person}{Nicholas Carlini} {and}
  \bibinfo{person}{David Wagner}.} \bibinfo{year}{2017}\natexlab{}.
\newblock \showarticletitle{Adversarial examples are not easily detected:
  Bypassing ten detection methods}. In \bibinfo{booktitle}{\emph{Proceedings of
  the 10th ACM Workshop on Artificial Intelligence and Security}}.
  \bibinfo{pages}{3--14}.
\newblock


\bibitem[\protect\citeauthoryear{Chen, Ravindranath, Deng, Bahl, and
  Balakrishnan}{Chen et~al\mbox{.}}{2015}]%
        {chen2015glimpse}
\bibfield{author}{\bibinfo{person}{Tiffany Yu-Han Chen}, \bibinfo{person}{Lenin
  Ravindranath}, \bibinfo{person}{Shuo Deng}, \bibinfo{person}{Paramvir Bahl},
  {and} \bibinfo{person}{Hari Balakrishnan}.} \bibinfo{year}{2015}\natexlab{}.
\newblock \showarticletitle{Glimpse: Continuous, real-time object recognition
  on mobile devices}. In \bibinfo{booktitle}{\emph{Proceedings of the 13th ACM
  Conference on Embedded Networked Sensor Systems}}. \bibinfo{pages}{155--168}.
\newblock


\bibitem[\protect\citeauthoryear{CNBC-Study}{CNBC-Study}{2019}]%
        {CNBC-STUDY}
\bibfield{author}{\bibinfo{person}{CNBC-Study}.}
  \bibinfo{year}{2019}\natexlab{}.
\newblock \bibinfo{title}{One billion surveillance cameras will be watching
  around the world in 2021, a new study says}.
\newblock
  \bibinfo{howpublished}{\href{https://www.cnbc.com/2019/12/06/one-billion-surveillance-cameras-will-be-watching-globally-in-2021.html}{cnbc\_study\_reports\_1blllion\_surveillancecamera\_by2021}}.
\newblock


\bibitem[\protect\citeauthoryear{CNET}{CNET}{2019}]%
        {CNET-5G}
\bibfield{author}{\bibinfo{person}{CNET}.} \bibinfo{year}{2019}\natexlab{}.
\newblock \bibinfo{title}{How 5G aims to end network latency}.
\newblock
  \bibinfo{howpublished}{\href{https://www.cnet.com/news/how-5g-aims-to-end-network-latency-response-time/}{CNET\_5G\_network\_latency\_time}}.
\newblock


\bibitem[\protect\citeauthoryear{cocoapi github}{cocoapi github}{[n.d.]}]%
        {pycoco}
\bibfield{author}{\bibinfo{person}{cocoapi github}.}
  \bibinfo{year}{}\natexlab{}.
\newblock \bibinfo{title}{pycocotools}.
\newblock
  \bibinfo{howpublished}{\url{https://github.com/cocodataset/cocoapi/tree/master/PythonAPI/pycocotools}}.
\newblock


\bibitem[\protect\citeauthoryear{Deng, Dong, Socher, Li, Li, and Fei-Fei}{Deng
  et~al\mbox{.}}{2009}]%
        {deng2009imagenet}
\bibfield{author}{\bibinfo{person}{Jia Deng}, \bibinfo{person}{Wei Dong},
  \bibinfo{person}{Richard Socher}, \bibinfo{person}{Li-Jia Li},
  \bibinfo{person}{Kai Li}, {and} \bibinfo{person}{Li Fei-Fei}.}
  \bibinfo{year}{2009}\natexlab{}.
\newblock \showarticletitle{Imagenet: A large-scale hierarchical image
  database}. In \bibinfo{booktitle}{\emph{2009 IEEE conference on computer
  vision and pattern recognition}}. IEEE, \bibinfo{pages}{248--255}.
\newblock


\bibitem[\protect\citeauthoryear{Deng, Han, and Ansari}{Deng
  et~al\mbox{.}}{2020}]%
        {deng2020fedvision}
\bibfield{author}{\bibinfo{person}{Yang Deng}, \bibinfo{person}{Tao Han}, {and}
  \bibinfo{person}{Nirwan Ansari}.} \bibinfo{year}{2020}\natexlab{}.
\newblock \showarticletitle{FedVision: Federated Video Analytics With Edge
  Computing}.
\newblock \bibinfo{journal}{\emph{IEEE Open Journal of the Computer Society}}
  \bibinfo{volume}{1} (\bibinfo{year}{2020}), \bibinfo{pages}{62--72}.
\newblock


\bibitem[\protect\citeauthoryear{Diamond, Sitzmann, Boyd, Wetzstein, and
  Heide}{Diamond et~al\mbox{.}}{2017}]%
        {DBLP:journals/corr/DiamondSBWH17}
\bibfield{author}{\bibinfo{person}{Steven Diamond}, \bibinfo{person}{Vincent
  Sitzmann}, \bibinfo{person}{Stephen~P. Boyd}, \bibinfo{person}{Gordon
  Wetzstein}, {and} \bibinfo{person}{Felix Heide}.}
  \bibinfo{year}{2017}\natexlab{}.
\newblock \showarticletitle{Dirty Pixels: Optimizing Image Classification
  Architectures for Raw Sensor Data}.
\newblock \bibinfo{journal}{\emph{CoRR}}  \bibinfo{volume}{abs/1701.06487}
  (\bibinfo{year}{2017}).
\newblock
\showeprint[arxiv]{1701.06487}
\urldef\tempurl%
\url{http://arxiv.org/abs/1701.06487}
\showURL{%
\tempurl}


\bibitem[\protect\citeauthoryear{Ding, Li, Cheng, Wang, Bian, and Jie}{Ding
  et~al\mbox{.}}{2020}]%
        {ding2020local}
\bibfield{author}{\bibinfo{person}{Xintao Ding}, \bibinfo{person}{Qingde Li},
  \bibinfo{person}{Yongqiang Cheng}, \bibinfo{person}{Jinbao Wang},
  \bibinfo{person}{Weixin Bian}, {and} \bibinfo{person}{Biao Jie}.}
  \bibinfo{year}{2020}\natexlab{}.
\newblock \showarticletitle{Local keypoint-based Faster R-CNN}.
\newblock \bibinfo{journal}{\emph{APPLIED INTELLIGENCE}}
  (\bibinfo{year}{2020}).
\newblock


\bibitem[\protect\citeauthoryear{Dodge and Karam}{Dodge and Karam}{2016}]%
        {dodge2016understanding}
\bibfield{author}{\bibinfo{person}{Samuel Dodge} {and} \bibinfo{person}{Lina
  Karam}.} \bibinfo{year}{2016}\natexlab{}.
\newblock \showarticletitle{Understanding how image quality affects deep neural
  networks}. In \bibinfo{booktitle}{\emph{2016 eighth international conference
  on quality of multimedia experience (QoMEX)}}. IEEE, \bibinfo{pages}{1--6}.
\newblock


\bibitem[\protect\citeauthoryear{Dodge and Karam}{Dodge and Karam}{2018}]%
        {dodge2018quality}
\bibfield{author}{\bibinfo{person}{Samuel~F Dodge} {and}
  \bibinfo{person}{Lina~J Karam}.} \bibinfo{year}{2018}\natexlab{}.
\newblock \showarticletitle{Quality robust mixtures of deep neural networks}.
\newblock \bibinfo{journal}{\emph{IEEE Transactions on Image Processing}}
  \bibinfo{volume}{27}, \bibinfo{number}{11} (\bibinfo{year}{2018}),
  \bibinfo{pages}{5553--5562}.
\newblock


\bibitem[\protect\citeauthoryear{facefirst}{facefirst}{2019}]%
        {sm1}
\bibfield{author}{\bibinfo{person}{facefirst}.}
  \bibinfo{year}{2019}\natexlab{}.
\newblock \bibinfo{title}{What does the future store look like?}
\newblock
  \bibinfo{howpublished}{\url{https://www.facefirst.com/blog/ways-future-stores-will-use-face-recognition-to-power-more-convenient-checkout/}}.
\newblock


\bibitem[\protect\citeauthoryear{Ghadiyaram, Pan, Bovik, Moorthy, Panda, and
  Yang}{Ghadiyaram et~al\mbox{.}}{2018}]%
        {chak-video-distortions1}
\bibfield{author}{\bibinfo{person}{D. Ghadiyaram}, \bibinfo{person}{J. Pan},
  \bibinfo{person}{A.~C. Bovik}, \bibinfo{person}{A.~K. Moorthy},
  \bibinfo{person}{P. Panda}, {and} \bibinfo{person}{K.~C. Yang}.}
  \bibinfo{year}{2018}\natexlab{}.
\newblock \showarticletitle{In-capture mobile video distortions: a study of
  subjective behavior and objective algorithms}.
\newblock \bibinfo{journal}{\emph{IEEE Transactions on Circuits and Systems for
  Video Technology}} \bibinfo{volume}{28}, \bibinfo{number}{9}
  (\bibinfo{year}{2018}), \bibinfo{pages}{2061-- 2077}.
\newblock


\bibitem[\protect\citeauthoryear{Goodfellow, Shlens, and Szegedy}{Goodfellow
  et~al\mbox{.}}{2014}]%
        {goodfellow2014explaining}
\bibfield{author}{\bibinfo{person}{Ian~J Goodfellow}, \bibinfo{person}{Jonathon
  Shlens}, {and} \bibinfo{person}{Christian Szegedy}.}
  \bibinfo{year}{2014}\natexlab{}.
\newblock \showarticletitle{Explaining and harnessing adversarial examples}.
\newblock \bibinfo{journal}{\emph{arXiv preprint arXiv:1412.6572}}
  (\bibinfo{year}{2014}).
\newblock


\bibitem[\protect\citeauthoryear{Gu, Wang, Kuen, Ma, Shahroudy, Shuai, Liu,
  Wang, Wang, Cai, et~al\mbox{.}}{Gu et~al\mbox{.}}{2018}]%
        {gu2018recent}
\bibfield{author}{\bibinfo{person}{Jiuxiang Gu}, \bibinfo{person}{Zhenhua
  Wang}, \bibinfo{person}{Jason Kuen}, \bibinfo{person}{Lianyang Ma},
  \bibinfo{person}{Amir Shahroudy}, \bibinfo{person}{Bing Shuai},
  \bibinfo{person}{Ting Liu}, \bibinfo{person}{Xingxing Wang},
  \bibinfo{person}{Gang Wang}, \bibinfo{person}{Jianfei Cai}, {et~al\mbox{.}}}
  \bibinfo{year}{2018}\natexlab{}.
\newblock \showarticletitle{Recent advances in convolutional neural networks}.
\newblock \bibinfo{journal}{\emph{Pattern Recognition}}  \bibinfo{volume}{77}
  (\bibinfo{year}{2018}), \bibinfo{pages}{354--377}.
\newblock


\bibitem[\protect\citeauthoryear{Guo and Zhang}{Guo and Zhang}{2019}]%
        {guo2019survey}
\bibfield{author}{\bibinfo{person}{Guodong Guo} {and} \bibinfo{person}{Na
  Zhang}.} \bibinfo{year}{2019}\natexlab{}.
\newblock \showarticletitle{A survey on deep learning based face recognition}.
\newblock \bibinfo{journal}{\emph{Computer Vision and Image Understanding}}
  \bibinfo{volume}{189} (\bibinfo{year}{2019}), \bibinfo{pages}{102805}.
\newblock


\bibitem[\protect\citeauthoryear{H264}{H264}{[n.d.]}]%
        {264}
\bibfield{author}{\bibinfo{person}{H264}.} \bibinfo{year}{}\natexlab{}.
\newblock \bibinfo{title}{H.264 Video Encoding}.
\newblock
  \bibinfo{howpublished}{\url{https://en.wikipedia.org/wiki/Advanced_Video_Coding}}.
\newblock


\bibitem[\protect\citeauthoryear{Haris, Shakhnarovich, and Ukita}{Haris
  et~al\mbox{.}}{2018}]%
        {DBLP:journals/corr/abs-1803-11316}
\bibfield{author}{\bibinfo{person}{Muhammad Haris}, \bibinfo{person}{Greg
  Shakhnarovich}, {and} \bibinfo{person}{Norimichi Ukita}.}
  \bibinfo{year}{2018}\natexlab{}.
\newblock \showarticletitle{Task-Driven Super Resolution: Object Detection in
  Low-resolution Images}.
\newblock \bibinfo{journal}{\emph{CoRR}}  \bibinfo{volume}{abs/1803.11316}
  (\bibinfo{year}{2018}).
\newblock
\showeprint[arxiv]{1803.11316}
\urldef\tempurl%
\url{http://arxiv.org/abs/1803.11316}
\showURL{%
\tempurl}


\bibitem[\protect\citeauthoryear{He, Gkioxari, Doll{\'a}r, and Girshick}{He
  et~al\mbox{.}}{2017}]%
        {he2017mask}
\bibfield{author}{\bibinfo{person}{Kaiming He}, \bibinfo{person}{Georgia
  Gkioxari}, \bibinfo{person}{Piotr Doll{\'a}r}, {and} \bibinfo{person}{Ross
  Girshick}.} \bibinfo{year}{2017}\natexlab{}.
\newblock \showarticletitle{Mask r-cnn}. In
  \bibinfo{booktitle}{\emph{Proceedings of the IEEE international conference on
  computer vision}}. \bibinfo{pages}{2961--2969}.
\newblock


\bibitem[\protect\citeauthoryear{He, Zhang, Ren, and Sun}{He
  et~al\mbox{.}}{2016}]%
        {he2016deep}
\bibfield{author}{\bibinfo{person}{Kaiming He}, \bibinfo{person}{Xiangyu
  Zhang}, \bibinfo{person}{Shaoqing Ren}, {and} \bibinfo{person}{Jian Sun}.}
  \bibinfo{year}{2016}\natexlab{}.
\newblock \showarticletitle{Deep residual learning for image recognition}. In
  \bibinfo{booktitle}{\emph{Proceedings of the IEEE conference on computer
  vision and pattern recognition}}. \bibinfo{pages}{770--778}.
\newblock


\bibitem[\protect\citeauthoryear{Howard, Sandler, Chu, Chen, Chen, Tan, Wang,
  Zhu, Pang, Vasudevan, et~al\mbox{.}}{Howard et~al\mbox{.}}{2019}]%
        {howard2019searching}
\bibfield{author}{\bibinfo{person}{Andrew Howard}, \bibinfo{person}{Mark
  Sandler}, \bibinfo{person}{Grace Chu}, \bibinfo{person}{Liang-Chieh Chen},
  \bibinfo{person}{Bo Chen}, \bibinfo{person}{Mingxing Tan},
  \bibinfo{person}{Weijun Wang}, \bibinfo{person}{Yukun Zhu},
  \bibinfo{person}{Ruoming Pang}, \bibinfo{person}{Vijay Vasudevan},
  {et~al\mbox{.}}} \bibinfo{year}{2019}\natexlab{}.
\newblock \showarticletitle{Searching for mobilenetv3}. In
  \bibinfo{booktitle}{\emph{Proceedings of the IEEE/CVF International
  Conference on Computer Vision}}. \bibinfo{pages}{1314--1324}.
\newblock


\bibitem[\protect\citeauthoryear{Jain, Zhang, Zhou, Ananthanarayanan, Jiang,
  Shu, Bahl, and Gonzalez}{Jain et~al\mbox{.}}{2020}]%
        {jainspatula}
\bibfield{author}{\bibinfo{person}{Samvit Jain}, \bibinfo{person}{Xun Zhang},
  \bibinfo{person}{Yuhao Zhou}, \bibinfo{person}{Ganesh Ananthanarayanan},
  \bibinfo{person}{Junchen Jiang}, \bibinfo{person}{Yuanchao Shu},
  \bibinfo{person}{Paramvir Bahl}, {and} \bibinfo{person}{Joseph Gonzalez}.}
  \bibinfo{year}{2020}\natexlab{}.
\newblock \showarticletitle{Spatula: Efficient cross-camera video analytics on
  large camera networks}.
\newblock  (\bibinfo{year}{2020}).
\newblock


\bibitem[\protect\citeauthoryear{Jiang, Ananthanarayanan, Bodik, Sen, and
  Stoica}{Jiang et~al\mbox{.}}{2018}]%
        {jiang2018chameleon}
\bibfield{author}{\bibinfo{person}{Junchen Jiang}, \bibinfo{person}{Ganesh
  Ananthanarayanan}, \bibinfo{person}{Peter Bodik}, \bibinfo{person}{Siddhartha
  Sen}, {and} \bibinfo{person}{Ion Stoica}.} \bibinfo{year}{2018}\natexlab{}.
\newblock \showarticletitle{Chameleon: scalable adaptation of video analytics}.
  In \bibinfo{booktitle}{\emph{Proceedings of the 2018 Conference of the ACM
  Special Interest Group on Data Communication}}. \bibinfo{pages}{253--266}.
\newblock


\bibitem[\protect\citeauthoryear{Jin, Wu, Li, Zhang, Chi, Peng, Ge, Zhao, and
  Li}{Jin et~al\mbox{.}}{2018}]%
        {jin2018ilgnet}
\bibfield{author}{\bibinfo{person}{Xin Jin}, \bibinfo{person}{Le Wu},
  \bibinfo{person}{Xiaodong Li}, \bibinfo{person}{Xiaokun Zhang},
  \bibinfo{person}{Jingying Chi}, \bibinfo{person}{Siwei Peng},
  \bibinfo{person}{Shiming Ge}, \bibinfo{person}{Geng Zhao}, {and}
  \bibinfo{person}{Shuying Li}.} \bibinfo{year}{2018}\natexlab{}.
\newblock \showarticletitle{ILGNet: inception modules with connected local and
  global features for efficient image aesthetic quality classification using
  domain adaptation}.
\newblock \bibinfo{journal}{\emph{IET Computer Vision}} \bibinfo{volume}{13},
  \bibinfo{number}{2} (\bibinfo{year}{2018}), \bibinfo{pages}{206--212}.
\newblock


\bibitem[\protect\citeauthoryear{Jordon}{Jordon}{[n.d.]}]%
        {CNN}
\bibfield{author}{\bibinfo{person}{Jeremy Jordon}.}
  \bibinfo{year}{}\natexlab{}.
\newblock \bibinfo{title}{Convolution Neural Network}.
\newblock
  \bibinfo{howpublished}{\url{https://www.jeremyjordan.me/convolutional-neural-networks/}}.
\newblock


\bibitem[\protect\citeauthoryear{JS}{JS}{[n.d.]}]%
        {JS}
\bibfield{author}{\bibinfo{person}{JS}.} \bibinfo{year}{}\natexlab{}.
\newblock \bibinfo{title}{Jensen\_Shannon Divegence}.
\newblock
  \bibinfo{howpublished}{\url{https://en.wikipedia.org/wiki/Jensen\%E2\%80\%93Shannon_divergence}}.
\newblock


\bibitem[\protect\citeauthoryear{Kang, Ye, Li, and Doermann}{Kang
  et~al\mbox{.}}{2014}]%
        {kang2014convolutional}
\bibfield{author}{\bibinfo{person}{Le Kang}, \bibinfo{person}{Peng Ye},
  \bibinfo{person}{Yi Li}, {and} \bibinfo{person}{David Doermann}.}
  \bibinfo{year}{2014}\natexlab{}.
\newblock \showarticletitle{Convolutional neural networks for no-reference
  image quality assessment}. In \bibinfo{booktitle}{\emph{Proceedings of the
  IEEE conference on computer vision and pattern recognition}}.
  \bibinfo{pages}{1733--1740}.
\newblock


\bibitem[\protect\citeauthoryear{Kingma and Ba}{Kingma and Ba}{2014}]%
        {kingma2014adam}
\bibfield{author}{\bibinfo{person}{Diederik~P Kingma} {and}
  \bibinfo{person}{Jimmy Ba}.} \bibinfo{year}{2014}\natexlab{}.
\newblock \showarticletitle{Adam: A method for stochastic optimization}.
\newblock \bibinfo{journal}{\emph{arXiv preprint arXiv:1412.6980}}
  (\bibinfo{year}{2014}).
\newblock


\bibitem[\protect\citeauthoryear{KL}{KL}{[n.d.]}]%
        {KL}
\bibfield{author}{\bibinfo{person}{KL}.} \bibinfo{year}{}\natexlab{}.
\newblock \bibinfo{title}{Kullback-Leibler Divegence}.
\newblock
  \bibinfo{howpublished}{\url{https://en.wikipedia.org/wiki/Kullback\%E2\%80\%93Leibler_divergence}}.
\newblock


\bibitem[\protect\citeauthoryear{Kortli, Jridi, Al~Falou, and Atri}{Kortli
  et~al\mbox{.}}{2020}]%
        {kortli2020face}
\bibfield{author}{\bibinfo{person}{Yassin Kortli}, \bibinfo{person}{Maher
  Jridi}, \bibinfo{person}{Ayman Al~Falou}, {and} \bibinfo{person}{Mohamed
  Atri}.} \bibinfo{year}{2020}\natexlab{}.
\newblock \showarticletitle{Face recognition systems: A Survey}.
\newblock \bibinfo{journal}{\emph{Sensors}} \bibinfo{volume}{20},
  \bibinfo{number}{2} (\bibinfo{year}{2020}), \bibinfo{pages}{342}.
\newblock


\bibitem[\protect\citeauthoryear{Krizhevsky, Sutskever, and Hinton}{Krizhevsky
  et~al\mbox{.}}{2012}]%
        {krizhevsky2012imagenet}
\bibfield{author}{\bibinfo{person}{Alex Krizhevsky}, \bibinfo{person}{Ilya
  Sutskever}, {and} \bibinfo{person}{Geoffrey~E Hinton}.}
  \bibinfo{year}{2012}\natexlab{}.
\newblock \showarticletitle{Imagenet classification with deep convolutional
  neural networks}. In \bibinfo{booktitle}{\emph{Advances in neural information
  processing systems}}. \bibinfo{pages}{1097--1105}.
\newblock


\bibitem[\protect\citeauthoryear{LeCun et~al\mbox{.}}{LeCun
  et~al\mbox{.}}{2015}]%
        {lecun2015lenet}
\bibfield{author}{\bibinfo{person}{Yann LeCun} {et~al\mbox{.}}}
  \bibinfo{year}{2015}\natexlab{}.
\newblock \showarticletitle{LeNet-5, convolutional neural networks}.
\newblock \bibinfo{journal}{\emph{URL: http://yann.lecun.com/exdb/lenet}}
  \bibinfo{volume}{20}, \bibinfo{number}{5} (\bibinfo{year}{2015}),
  \bibinfo{pages}{14}.
\newblock


\bibitem[\protect\citeauthoryear{Li, Padmanabhan, Zhao, Wang, Xu, and
  Netravali}{Li et~al\mbox{.}}{2020}]%
        {li2020reducto}
\bibfield{author}{\bibinfo{person}{Yuanqi Li}, \bibinfo{person}{Arthi
  Padmanabhan}, \bibinfo{person}{Pengzhan Zhao}, \bibinfo{person}{Yufei Wang},
  \bibinfo{person}{Guoqing~Harry Xu}, {and} \bibinfo{person}{Ravi Netravali}.}
  \bibinfo{year}{2020}\natexlab{}.
\newblock \showarticletitle{Reducto: On-Camera Filtering for Resource-Efficient
  Real-Time Video Analytics}. In \bibinfo{booktitle}{\emph{Proceedings of the
  Annual conference of the ACM Special Interest Group on Data Communication on
  the applications, technologies, architectures, and protocols for computer
  communication}}. \bibinfo{pages}{359--376}.
\newblock


\bibitem[\protect\citeauthoryear{Liang, Shenoy, and Irwin}{Liang
  et~al\mbox{.}}{2020}]%
        {liang2020ai}
\bibfield{author}{\bibinfo{person}{Qianlin Liang}, \bibinfo{person}{Prashant
  Shenoy}, {and} \bibinfo{person}{David Irwin}.}
  \bibinfo{year}{2020}\natexlab{}.
\newblock \showarticletitle{AI on the Edge: Rethinking AI-based IoT
  Applications Using Specialized Edge Architectures}.
\newblock \bibinfo{journal}{\emph{arXiv preprint arXiv:2003.12488}}
  (\bibinfo{year}{2020}).
\newblock


\bibitem[\protect\citeauthoryear{Lin, Maire, Belongie, Hays, Perona, Ramanan,
  Doll{\'a}r, and Zitnick}{Lin et~al\mbox{.}}{2014}]%
        {lin2014microsoft}
\bibfield{author}{\bibinfo{person}{Tsung-Yi Lin}, \bibinfo{person}{Michael
  Maire}, \bibinfo{person}{Serge Belongie}, \bibinfo{person}{James Hays},
  \bibinfo{person}{Pietro Perona}, \bibinfo{person}{Deva Ramanan},
  \bibinfo{person}{Piotr Doll{\'a}r}, {and} \bibinfo{person}{C~Lawrence
  Zitnick}.} \bibinfo{year}{2014}\natexlab{}.
\newblock \showarticletitle{Microsoft coco: Common objects in context}. In
  \bibinfo{booktitle}{\emph{European conference on computer vision}}. Springer,
  \bibinfo{pages}{740--755}.
\newblock


\bibitem[\protect\citeauthoryear{Liu, Li, and Gruteser}{Liu
  et~al\mbox{.}}{2019}]%
        {liu2019edge}
\bibfield{author}{\bibinfo{person}{Luyang Liu}, \bibinfo{person}{Hongyu Li},
  {and} \bibinfo{person}{Marco Gruteser}.} \bibinfo{year}{2019}\natexlab{}.
\newblock \showarticletitle{Edge assisted real-time object detection for mobile
  augmented reality}. In \bibinfo{booktitle}{\emph{The 25th Annual
  International Conference on Mobile Computing and Networking}}.
  \bibinfo{pages}{1--16}.
\newblock


\bibitem[\protect\citeauthoryear{Liu, Luo, Wang, and Tang}{Liu
  et~al\mbox{.}}{2015}]%
        {liu2015faceattributes}
\bibfield{author}{\bibinfo{person}{Ziwei Liu}, \bibinfo{person}{Ping Luo},
  \bibinfo{person}{Xiaogang Wang}, {and} \bibinfo{person}{Xiaoou Tang}.}
  \bibinfo{year}{2015}\natexlab{}.
\newblock \showarticletitle{Deep Learning Face Attributes in the Wild}. In
  \bibinfo{booktitle}{\emph{Proceedings of International Conference on Computer
  Vision (ICCV)}}.
\newblock


\bibitem[\protect\citeauthoryear{Market}{Market}{2019}]%
        {VIDEO-ANALYTICS-MARKET}
\bibfield{author}{\bibinfo{person}{Analytics Market}.}
  \bibinfo{year}{2019}\natexlab{}.
\newblock \bibinfo{title}{Video Analytics Market Statistics: 2027}.
\newblock
  \bibinfo{howpublished}{\url{https://www.alliedmarketresearch.com/video-analytics-market}}.
\newblock


\bibitem[\protect\citeauthoryear{Mittal, Moorthy, and Bovik}{Mittal
  et~al\mbox{.}}{2012}]%
        {brisque_mittal2012no}
\bibfield{author}{\bibinfo{person}{Anish Mittal},
  \bibinfo{person}{Anush~Krishna Moorthy}, {and} \bibinfo{person}{Alan~Conrad
  Bovik}.} \bibinfo{year}{2012}\natexlab{}.
\newblock \showarticletitle{No-reference image quality assessment in the
  spatial domain}.
\newblock \bibinfo{journal}{\emph{IEEE Transactions on image processing}}
  \bibinfo{volume}{21}, \bibinfo{number}{12} (\bibinfo{year}{2012}),
  \bibinfo{pages}{4695--4708}.
\newblock


\bibitem[\protect\citeauthoryear{Moon, Chang, and Lee}{Moon
  et~al\mbox{.}}{2019}]%
        {moon2019posefix}
\bibfield{author}{\bibinfo{person}{Gyeongsik Moon}, \bibinfo{person}{Ju~Yong
  Chang}, {and} \bibinfo{person}{Kyoung~Mu Lee}.}
  \bibinfo{year}{2019}\natexlab{}.
\newblock \showarticletitle{Posefix: Model-agnostic general human pose
  refinement network}. In \bibinfo{booktitle}{\emph{Proceedings of the IEEE
  Conference on Computer Vision and Pattern Recognition}}.
  \bibinfo{pages}{7773--7781}.
\newblock


\bibitem[\protect\citeauthoryear{Moorthy and Bovik}{Moorthy and Bovik}{2010}]%
        {biqi_moorthy2010two}
\bibfield{author}{\bibinfo{person}{Anush~Krishna Moorthy} {and}
  \bibinfo{person}{Alan~Conrad Bovik}.} \bibinfo{year}{2010}\natexlab{}.
\newblock \showarticletitle{A two-step framework for constructing blind image
  quality indices}.
\newblock \bibinfo{journal}{\emph{IEEE Signal processing letters}}
  \bibinfo{volume}{17}, \bibinfo{number}{5} (\bibinfo{year}{2010}),
  \bibinfo{pages}{513--516}.
\newblock


\bibitem[\protect\citeauthoryear{Moorthy and Bovik}{Moorthy and Bovik}{2011}]%
        {divine_moorthy2011blind}
\bibfield{author}{\bibinfo{person}{Anush~Krishna Moorthy} {and}
  \bibinfo{person}{Alan~Conrad Bovik}.} \bibinfo{year}{2011}\natexlab{}.
\newblock \showarticletitle{Blind image quality assessment: From natural scene
  statistics to perceptual quality}.
\newblock \bibinfo{journal}{\emph{IEEE transactions on Image Processing}}
  \bibinfo{volume}{20}, \bibinfo{number}{12} (\bibinfo{year}{2011}),
  \bibinfo{pages}{3350--3364}.
\newblock


\bibitem[\protect\citeauthoryear{Murray, Marchesotti, and Perronnin}{Murray
  et~al\mbox{.}}{2012}]%
        {AVA}
\bibfield{author}{\bibinfo{person}{Naila Murray}, \bibinfo{person}{Luca
  Marchesotti}, {and} \bibinfo{person}{Florent Perronnin}.}
  \bibinfo{year}{2012}\natexlab{}.
\newblock \showarticletitle{AVA: A large-scale database for aesthetic visual
  analysis}. In \bibinfo{booktitle}{\emph{2012 IEEE Conference on Computer
  Vision and Pattern Recognition}}. IEEE, \bibinfo{pages}{2408--2415}.
\newblock


\bibitem[\protect\citeauthoryear{Ng and Winkler}{Ng and Winkler}{2014}]%
        {ng2014data}
\bibfield{author}{\bibinfo{person}{Hong-Wei Ng} {and} \bibinfo{person}{Stefan
  Winkler}.} \bibinfo{year}{2014}\natexlab{}.
\newblock \showarticletitle{A data-driven approach to cleaning large face
  datasets}. In \bibinfo{booktitle}{\emph{2014 IEEE international conference on
  image processing (ICIP)}}. IEEE, \bibinfo{pages}{343--347}.
\newblock


\bibitem[\protect\citeauthoryear{Pei, Huang, Zou, Zang, Zhang, and Wang}{Pei
  et~al\mbox{.}}{2018}]%
        {pei2018effects}
\bibfield{author}{\bibinfo{person}{Yanting Pei}, \bibinfo{person}{Yaping
  Huang}, \bibinfo{person}{Qi Zou}, \bibinfo{person}{Hao Zang},
  \bibinfo{person}{Xingyuan Zhang}, {and} \bibinfo{person}{Song Wang}.}
  \bibinfo{year}{2018}\natexlab{}.
\newblock \showarticletitle{Effects of image degradations to CNN-based image
  classification}.
\newblock \bibinfo{journal}{\emph{arXiv preprint arXiv:1810.05552}}
  (\bibinfo{year}{2018}).
\newblock


\bibitem[\protect\citeauthoryear{Ponomarenko, Ieremeiev, Lukin, Egiazarian,
  Jin, Astola, Vozel, Chehdi, Carli, Battisti, et~al\mbox{.}}{Ponomarenko
  et~al\mbox{.}}{2013}]%
        {TID2013}
\bibfield{author}{\bibinfo{person}{Nikolay Ponomarenko}, \bibinfo{person}{Oleg
  Ieremeiev}, \bibinfo{person}{Vladimir Lukin}, \bibinfo{person}{Karen
  Egiazarian}, \bibinfo{person}{Lina Jin}, \bibinfo{person}{Jaakko Astola},
  \bibinfo{person}{Benoit Vozel}, \bibinfo{person}{Kacem Chehdi},
  \bibinfo{person}{Marco Carli}, \bibinfo{person}{Federica Battisti},
  {et~al\mbox{.}}} \bibinfo{year}{2013}\natexlab{}.
\newblock \showarticletitle{Color image database TID2013: Peculiarities and
  preliminary results}. In \bibinfo{booktitle}{\emph{european workshop on
  visual information processing (EUVIP)}}. IEEE, \bibinfo{pages}{106--111}.
\newblock


\bibitem[\protect\citeauthoryear{pytorch}{pytorch}{[n.d.]}]%
        {pytorch_pretrained}
\bibfield{author}{\bibinfo{person}{pytorch}.} \bibinfo{year}{}\natexlab{}.
\newblock \bibinfo{title}{Pretrained Models}.
\newblock
  \bibinfo{howpublished}{\url{https://pytorch.org/docs/stable/torchvision/models.html}}.
\newblock


\bibitem[\protect\citeauthoryear{Qiao, Chen, and Yuille}{Qiao
  et~al\mbox{.}}{2020}]%
        {qiao2020detectors}
\bibfield{author}{\bibinfo{person}{Siyuan Qiao}, \bibinfo{person}{Liang-Chieh
  Chen}, {and} \bibinfo{person}{Alan Yuille}.} \bibinfo{year}{2020}\natexlab{}.
\newblock \showarticletitle{DetectoRS: Detecting Objects with Recursive Feature
  Pyramid and Switchable Atrous Convolution}.
\newblock \bibinfo{journal}{\emph{arXiv preprint arXiv:2006.02334}}
  (\bibinfo{year}{2020}).
\newblock


\bibitem[\protect\citeauthoryear{Qualcomm}{Qualcomm}{2019}]%
        {QUALCOMM-5G}
\bibfield{author}{\bibinfo{person}{Qualcomm}.} \bibinfo{year}{2019}\natexlab{}.
\newblock \bibinfo{title}{How 5G low latency improves your mobile experiences}.
\newblock
  \bibinfo{howpublished}{\href{https://www.qualcomm.com/news/onq/2019/05/13/how-5g-low-latency-improves-your-mobile-experiences/}{Qualcomm\_5G\_low-latency\_improves\_mobile\_experience}}.
\newblock


\bibitem[\protect\citeauthoryear{Ranjan, Bansal, Zheng, Xu, Gleason, Lu,
  Nanduri, Chen, Castillo, and Chellappa}{Ranjan et~al\mbox{.}}{2019}]%
        {ranjan2019fast}
\bibfield{author}{\bibinfo{person}{Rajeev Ranjan}, \bibinfo{person}{Ankan
  Bansal}, \bibinfo{person}{Jingxiao Zheng}, \bibinfo{person}{Hongyu Xu},
  \bibinfo{person}{Joshua Gleason}, \bibinfo{person}{Boyu Lu},
  \bibinfo{person}{Anirudh Nanduri}, \bibinfo{person}{Jun-Cheng Chen},
  \bibinfo{person}{Carlos~D Castillo}, {and} \bibinfo{person}{Rama Chellappa}.}
  \bibinfo{year}{2019}\natexlab{}.
\newblock \showarticletitle{A fast and accurate system for face detection,
  identification, and verification}.
\newblock \bibinfo{journal}{\emph{IEEE Transactions on Biometrics, Behavior,
  and Identity Science}} \bibinfo{volume}{1}, \bibinfo{number}{2}
  (\bibinfo{year}{2019}), \bibinfo{pages}{82--96}.
\newblock


\bibitem[\protect\citeauthoryear{Redmon, Divvala, Girshick, and Farhadi}{Redmon
  et~al\mbox{.}}{2016}]%
        {yoloredmon2016you}
\bibfield{author}{\bibinfo{person}{Joseph Redmon}, \bibinfo{person}{Santosh
  Divvala}, \bibinfo{person}{Ross Girshick}, {and} \bibinfo{person}{Ali
  Farhadi}.} \bibinfo{year}{2016}\natexlab{}.
\newblock \showarticletitle{You only look once: Unified, real-time object
  detection}. In \bibinfo{booktitle}{\emph{Proceedings of the IEEE conference
  on computer vision and pattern recognition}}. \bibinfo{pages}{779--788}.
\newblock


\bibitem[\protect\citeauthoryear{Ren, He, Girshick, and Sun}{Ren
  et~al\mbox{.}}{2015}]%
        {ren2015faster}
\bibfield{author}{\bibinfo{person}{Shaoqing Ren}, \bibinfo{person}{Kaiming He},
  \bibinfo{person}{Ross Girshick}, {and} \bibinfo{person}{Jian Sun}.}
  \bibinfo{year}{2015}\natexlab{}.
\newblock \showarticletitle{Faster r-cnn: Towards real-time object detection
  with region proposal networks}. In \bibinfo{booktitle}{\emph{Advances in
  neural information processing systems}}. \bibinfo{pages}{91--99}.
\newblock


\bibitem[\protect\citeauthoryear{retail-customer experience}{retail-customer
  experience}{2019}]%
        {sm2}
\bibfield{author}{\bibinfo{person}{retail-customer experience}.}
  \bibinfo{year}{2019}\natexlab{}.
\newblock \bibinfo{title}{Shopping centers quietly test facial recognition
  technology}.
\newblock
  \bibinfo{howpublished}{\url{https://www.retailcustomerexperience.com/news/shopping-centers-quietly-test-facial-recognition-technology/}}.
\newblock


\bibitem[\protect\citeauthoryear{Roussi}{Roussi}{2020}]%
        {r2}
\bibfield{author}{\bibinfo{person}{Antoaneta Roussi}.}
  \bibinfo{year}{2020}\natexlab{}.
\newblock \bibinfo{title}{Resisting the rise of facial recognition}.
\newblock
  \bibinfo{howpublished}{\url{https://www.nature.com/articles/d41586-020-03188-2}}.
\newblock


\bibitem[\protect\citeauthoryear{Roy, Ghosh, Bhattacharya, and Pal}{Roy
  et~al\mbox{.}}{2018}]%
        {roy2018effects}
\bibfield{author}{\bibinfo{person}{Prasun Roy}, \bibinfo{person}{Subhankar
  Ghosh}, \bibinfo{person}{Saumik Bhattacharya}, {and} \bibinfo{person}{Umapada
  Pal}.} \bibinfo{year}{2018}\natexlab{}.
\newblock \showarticletitle{Effects of degradations on deep neural network
  architectures}.
\newblock \bibinfo{journal}{\emph{arXiv preprint arXiv:1807.10108}}
  (\bibinfo{year}{2018}).
\newblock


\bibitem[\protect\citeauthoryear{Saad, Bovik, and Charrier}{Saad
  et~al\mbox{.}}{2012}]%
        {blinds2_saad2012blind}
\bibfield{author}{\bibinfo{person}{Michele~A Saad}, \bibinfo{person}{Alan~C
  Bovik}, {and} \bibinfo{person}{Christophe Charrier}.}
  \bibinfo{year}{2012}\natexlab{}.
\newblock \showarticletitle{Blind image quality assessment: A natural scene
  statistics approach in the DCT domain}.
\newblock \bibinfo{journal}{\emph{IEEE transactions on Image Processing}}
  \bibinfo{volume}{21}, \bibinfo{number}{8} (\bibinfo{year}{2012}),
  \bibinfo{pages}{3339--3352}.
\newblock


\bibitem[\protect\citeauthoryear{Sajjad, Nasir, Muhammad, Khan, Jan, Sangaiah,
  Elhoseny, and Baik}{Sajjad et~al\mbox{.}}{2020}]%
        {sajjad2020raspberry}
\bibfield{author}{\bibinfo{person}{Muhammad Sajjad}, \bibinfo{person}{Mansoor
  Nasir}, \bibinfo{person}{Khan Muhammad}, \bibinfo{person}{Siraj Khan},
  \bibinfo{person}{Zahoor Jan}, \bibinfo{person}{Arun~Kumar Sangaiah},
  \bibinfo{person}{Mohamed Elhoseny}, {and} \bibinfo{person}{Sung~Wook Baik}.}
  \bibinfo{year}{2020}\natexlab{}.
\newblock \showarticletitle{Raspberry Pi assisted face recognition framework
  for enhanced law-enforcement services in smart cities}.
\newblock \bibinfo{journal}{\emph{Future Generation Computer Systems}}
  \bibinfo{volume}{108} (\bibinfo{year}{2020}), \bibinfo{pages}{995--1007}.
\newblock


\bibitem[\protect\citeauthoryear{Schroff, Kalenichenko, and Philbin}{Schroff
  et~al\mbox{.}}{2015}]%
        {schroff2015facenet}
\bibfield{author}{\bibinfo{person}{Florian Schroff}, \bibinfo{person}{Dmitry
  Kalenichenko}, {and} \bibinfo{person}{James Philbin}.}
  \bibinfo{year}{2015}\natexlab{}.
\newblock \showarticletitle{Facenet: A unified embedding for face recognition
  and clustering}. In \bibinfo{booktitle}{\emph{Proceedings of the IEEE
  conference on computer vision and pattern recognition}}.
  \bibinfo{pages}{815--823}.
\newblock


\bibitem[\protect\citeauthoryear{Seshadrinathan, Soundarajan, Bovik, and
  Cormack}{Seshadrinathan et~al\mbox{.}}{2010}]%
        {chak-video-distortions2}
\bibfield{author}{\bibinfo{person}{K. Seshadrinathan}, \bibinfo{person}{R.
  Soundarajan}, \bibinfo{person}{A.~C. Bovik}, {and} \bibinfo{person}{L.~K.
  Cormack}.} \bibinfo{year}{2010}\natexlab{}.
\newblock \showarticletitle{Study of subjective and objective quality
  assessment of video}.
\newblock \bibinfo{journal}{\emph{IEEE Transactions on Image Processing}}
  \bibinfo{volume}{19}, \bibinfo{number}{6} (\bibinfo{year}{2010}),
  \bibinfo{pages}{1427--1441}.
\newblock


\bibitem[\protect\citeauthoryear{Sheikh}{Sheikh}{2005}]%
        {LIVE}
\bibfield{author}{\bibinfo{person}{Hamid~R Sheikh}.}
  \bibinfo{year}{2005}\natexlab{}.
\newblock \showarticletitle{LIVE image quality assessment database}.
\newblock \bibinfo{journal}{\emph{http://live. ece. utexas.
  edu/research/quality}} (\bibinfo{year}{2005}).
\newblock


\bibitem[\protect\citeauthoryear{Simonyan and Zisserman}{Simonyan and
  Zisserman}{2014}]%
        {vgg19simonyan2014very}
\bibfield{author}{\bibinfo{person}{Karen Simonyan} {and}
  \bibinfo{person}{Andrew Zisserman}.} \bibinfo{year}{2014}\natexlab{}.
\newblock \showarticletitle{Very deep convolutional networks for large-scale
  image recognition}.
\newblock \bibinfo{journal}{\emph{arXiv preprint arXiv:1409.1556}}
  (\bibinfo{year}{2014}).
\newblock


\bibitem[\protect\citeauthoryear{Szegedy, Liu, Jia, Sermanet, Reed, Anguelov,
  Erhan, Vanhoucke, and Rabinovich}{Szegedy et~al\mbox{.}}{2015}]%
        {szegedy2015going}
\bibfield{author}{\bibinfo{person}{Christian Szegedy}, \bibinfo{person}{Wei
  Liu}, \bibinfo{person}{Yangqing Jia}, \bibinfo{person}{Pierre Sermanet},
  \bibinfo{person}{Scott Reed}, \bibinfo{person}{Dragomir Anguelov},
  \bibinfo{person}{Dumitru Erhan}, \bibinfo{person}{Vincent Vanhoucke}, {and}
  \bibinfo{person}{Andrew Rabinovich}.} \bibinfo{year}{2015}\natexlab{}.
\newblock \showarticletitle{Going deeper with convolutions}. In
  \bibinfo{booktitle}{\emph{Proceedings of the IEEE conference on computer
  vision and pattern recognition}}. \bibinfo{pages}{1--9}.
\newblock


\bibitem[\protect\citeauthoryear{Szegedy, Vanhoucke, Ioffe, Shlens, and
  Wojna}{Szegedy et~al\mbox{.}}{2016}]%
        {szegedy2016rethinking}
\bibfield{author}{\bibinfo{person}{Christian Szegedy}, \bibinfo{person}{Vincent
  Vanhoucke}, \bibinfo{person}{Sergey Ioffe}, \bibinfo{person}{Jon Shlens},
  {and} \bibinfo{person}{Zbigniew Wojna}.} \bibinfo{year}{2016}\natexlab{}.
\newblock \showarticletitle{Rethinking the inception architecture for computer
  vision}. In \bibinfo{booktitle}{\emph{Proceedings of the IEEE conference on
  computer vision and pattern recognition}}. \bibinfo{pages}{2818--2826}.
\newblock


\bibitem[\protect\citeauthoryear{Tadros, Cullen, Greene, and Cooper}{Tadros
  et~al\mbox{.}}{2019}]%
        {tadros2019assessing}
\bibfield{author}{\bibinfo{person}{Timothy Tadros}, \bibinfo{person}{Nicholas~C
  Cullen}, \bibinfo{person}{Michelle~R Greene}, {and} \bibinfo{person}{Emily~A
  Cooper}.} \bibinfo{year}{2019}\natexlab{}.
\newblock \showarticletitle{Assessing Neural Network Scene Classification from
  Degraded Images}.
\newblock \bibinfo{journal}{\emph{ACM Transactions on Applied Perception
  (TAP)}} \bibinfo{volume}{16}, \bibinfo{number}{4} (\bibinfo{year}{2019}),
  \bibinfo{pages}{1--20}.
\newblock


\bibitem[\protect\citeauthoryear{Talebi and Milanfar}{Talebi and
  Milanfar}{2018}]%
        {talebi2018nima}
\bibfield{author}{\bibinfo{person}{Hossein Talebi} {and}
  \bibinfo{person}{Peyman Milanfar}.} \bibinfo{year}{2018}\natexlab{}.
\newblock \showarticletitle{NIMA: Neural image assessment}.
\newblock \bibinfo{journal}{\emph{IEEE Transactions on Image Processing}}
  \bibinfo{volume}{27}, \bibinfo{number}{8} (\bibinfo{year}{2018}),
  \bibinfo{pages}{3998--4011}.
\newblock


\bibitem[\protect\citeauthoryear{Tao}{Tao}{2018}]%
        {r1}
\bibfield{author}{\bibinfo{person}{Li Tao}.} \bibinfo{year}{2018}\natexlab{}.
\newblock \bibinfo{title}{Shenzhen police can now identify drivers using facial
  recognition surveillance cameras}.
\newblock
  \bibinfo{howpublished}{\url{https://www.scmp.com/tech/china-tech/article/2143137/shenzhen-police-can-now-identify-drivers-using-facial-recognition/}}.
\newblock


\bibitem[\protect\citeauthoryear{Vasiljevic, Chakrabarti, and
  Shakhnarovich}{Vasiljevic et~al\mbox{.}}{2016}]%
        {DBLP:journals/corr/VasiljevicCS16}
\bibfield{author}{\bibinfo{person}{Igor Vasiljevic}, \bibinfo{person}{Ayan
  Chakrabarti}, {and} \bibinfo{person}{Gregory Shakhnarovich}.}
  \bibinfo{year}{2016}\natexlab{}.
\newblock \showarticletitle{Examining the Impact of Blur on Recognition by
  Convolutional Networks}.
\newblock \bibinfo{journal}{\emph{CoRR}}  \bibinfo{volume}{abs/1611.05760}
  (\bibinfo{year}{2016}).
\newblock
\showeprint[arxiv]{1611.05760}
\urldef\tempurl%
\url{http://arxiv.org/abs/1611.05760}
\showURL{%
\tempurl}


\bibitem[\protect\citeauthoryear{Verge}{Verge}{2020}]%
        {ap1}
\bibfield{author}{\bibinfo{person}{Verge}.} \bibinfo{year}{2020}\natexlab{}.
\newblock \bibinfo{title}{Major expansion of facial recognition authority at
  airports}.
\newblock
  \bibinfo{howpublished}{\url{https://www.theverge.com/2020/12/18/22188526/airport-facial-recognition-us-customs-biometric-exit-expansion}}.
\newblock


\bibitem[\protect\citeauthoryear{VP9}{VP9}{[n.d.]}]%
        {VP9}
\bibfield{author}{\bibinfo{person}{VP9}.} \bibinfo{year}{}\natexlab{}.
\newblock \bibinfo{title}{VP9 Video Codec}.
\newblock \bibinfo{howpublished}{\url{https://en.wikipedia.org/wiki/VP9}}.
\newblock


\bibitem[\protect\citeauthoryear{Wall-Street-Journal}{Wall-Street-Journal}{2020}]%
        {ap2}
\bibfield{author}{\bibinfo{person}{Wall-Street-Journal}.}
  \bibinfo{year}{2020}\natexlab{}.
\newblock \bibinfo{title}{Are You Ready for Facial Recognition at the Airport?}
\newblock
  \bibinfo{howpublished}{\url{https://www.wsj.com/articles/are-you-ready-for-facial-recognition-at-the-airport-11565775008}}.
\newblock


\bibitem[\protect\citeauthoryear{Wang, Amos, Das, Pillai, Sadeh, and
  Satyanarayanan}{Wang et~al\mbox{.}}{2017}]%
        {wang2017scalable}
\bibfield{author}{\bibinfo{person}{Junjue Wang}, \bibinfo{person}{Brandon
  Amos}, \bibinfo{person}{Anupam Das}, \bibinfo{person}{Padmanabhan Pillai},
  \bibinfo{person}{Norman Sadeh}, {and} \bibinfo{person}{Mahadev
  Satyanarayanan}.} \bibinfo{year}{2017}\natexlab{}.
\newblock \showarticletitle{A scalable and privacy-aware IoT service for live
  video analytics}. In \bibinfo{booktitle}{\emph{Proceedings of the 8th ACM on
  Multimedia Systems Conference}}. \bibinfo{pages}{38--49}.
\newblock


\bibitem[\protect\citeauthoryear{Xue, Mou, Zhang, Bovik, and Feng}{Xue
  et~al\mbox{.}}{2014}]%
        {gmlog_xue2014blind}
\bibfield{author}{\bibinfo{person}{Wufeng Xue}, \bibinfo{person}{Xuanqin Mou},
  \bibinfo{person}{Lei Zhang}, \bibinfo{person}{Alan~C Bovik}, {and}
  \bibinfo{person}{Xiangchu Feng}.} \bibinfo{year}{2014}\natexlab{}.
\newblock \showarticletitle{Blind image quality assessment using joint
  statistics of gradient magnitude and Laplacian features}.
\newblock \bibinfo{journal}{\emph{IEEE Transactions on Image Processing}}
  \bibinfo{volume}{23}, \bibinfo{number}{11} (\bibinfo{year}{2014}),
  \bibinfo{pages}{4850--4862}.
\newblock


\bibitem[\protect\citeauthoryear{Yang, Ma, and Yang}{Yang
  et~al\mbox{.}}{2014}]%
        {yang2014single}
\bibfield{author}{\bibinfo{person}{Chih-Yuan Yang}, \bibinfo{person}{Chao Ma},
  {and} \bibinfo{person}{Ming-Hsuan Yang}.} \bibinfo{year}{2014}\natexlab{}.
\newblock \showarticletitle{Single-image super-resolution: A benchmark}. In
  \bibinfo{booktitle}{\emph{European Conference on Computer Vision}}. Springer,
  \bibinfo{pages}{372--386}.
\newblock


\bibitem[\protect\citeauthoryear{Ye, Kumar, Kang, and Doermann}{Ye
  et~al\mbox{.}}{2012}]%
        {cornia_ye2012unsupervised}
\bibfield{author}{\bibinfo{person}{Peng Ye}, \bibinfo{person}{Jayant Kumar},
  \bibinfo{person}{Le Kang}, {and} \bibinfo{person}{David Doermann}.}
  \bibinfo{year}{2012}\natexlab{}.
\newblock \showarticletitle{Unsupervised feature learning framework for
  no-reference image quality assessment}. In \bibinfo{booktitle}{\emph{2012
  IEEE conference on computer vision and pattern recognition}}. IEEE,
  \bibinfo{pages}{1098--1105}.
\newblock


\bibitem[\protect\citeauthoryear{Yi, Choi, and Lee}{Yi et~al\mbox{.}}{2020}]%
        {yi2020eagleeye}
\bibfield{author}{\bibinfo{person}{Juheon Yi}, \bibinfo{person}{Sunghyun Choi},
  {and} \bibinfo{person}{Youngki Lee}.} \bibinfo{year}{2020}\natexlab{}.
\newblock \showarticletitle{EagleEye: wearable camera-based person
  identification in crowded urban spaces}. In
  \bibinfo{booktitle}{\emph{Proceedings of the 26th Annual International
  Conference on Mobile Computing and Networking}}. \bibinfo{pages}{1--14}.
\newblock


\bibitem[\protect\citeauthoryear{Zhang, Jin, Ratnasamy, Wawrzynek, and
  Lee}{Zhang et~al\mbox{.}}{2018}]%
        {zhang2018awstream}
\bibfield{author}{\bibinfo{person}{Ben Zhang}, \bibinfo{person}{Xin Jin},
  \bibinfo{person}{Sylvia Ratnasamy}, \bibinfo{person}{John Wawrzynek}, {and}
  \bibinfo{person}{Edward~A Lee}.} \bibinfo{year}{2018}\natexlab{}.
\newblock \showarticletitle{Awstream: Adaptive wide-area streaming analytics}.
  In \bibinfo{booktitle}{\emph{Proceedings of the 2018 Conference of the ACM
  Special Interest Group on Data Communication}}. \bibinfo{pages}{236--252}.
\newblock


\bibitem[\protect\citeauthoryear{Zhang, Ananthanarayanan, Bodik, Philipose,
  Bahl, and Freedman}{Zhang et~al\mbox{.}}{2017}]%
        {zhang2017live}
\bibfield{author}{\bibinfo{person}{Haoyu Zhang}, \bibinfo{person}{Ganesh
  Ananthanarayanan}, \bibinfo{person}{Peter Bodik}, \bibinfo{person}{Matthai
  Philipose}, \bibinfo{person}{Paramvir Bahl}, {and} \bibinfo{person}{Michael~J
  Freedman}.} \bibinfo{year}{2017}\natexlab{}.
\newblock \showarticletitle{Live video analytics at scale with approximation
  and delay-tolerance}. In \bibinfo{booktitle}{\emph{14th $\{$USENIX$\}$
  Symposium on Networked Systems Design and Implementation ($\{$NSDI$\}$ 17)}}.
  \bibinfo{pages}{377--392}.
\newblock


\bibitem[\protect\citeauthoryear{Zheng, Song, Leung, and Goodfellow}{Zheng
  et~al\mbox{.}}{2016}]%
        {zheng2016improving}
\bibfield{author}{\bibinfo{person}{Stephan Zheng}, \bibinfo{person}{Yang Song},
  \bibinfo{person}{Thomas Leung}, {and} \bibinfo{person}{Ian Goodfellow}.}
  \bibinfo{year}{2016}\natexlab{}.
\newblock \bibinfo{title}{Improving the Robustness of Deep Neural Networks via
  Stability Training}.
\newblock
\newblock
\showeprint[arxiv]{1604.04326}~[cs.CV]


\bibitem[\protect\citeauthoryear{Zhou, Song, and Cheung}{Zhou
  et~al\mbox{.}}{2017}]%
        {zhou2017classification}
\bibfield{author}{\bibinfo{person}{Yiren Zhou}, \bibinfo{person}{Sibo Song},
  {and} \bibinfo{person}{Ngai-Man Cheung}.} \bibinfo{year}{2017}\natexlab{}.
\newblock \showarticletitle{On classification of distorted images with deep
  convolutional neural networks}. In \bibinfo{booktitle}{\emph{2017 IEEE
  International Conference on Acoustics, Speech and Signal Processing
  (ICASSP)}}. IEEE, \bibinfo{pages}{1213--1217}.
\newblock


\end{thebibliography}
\bibliographystyle{ACM-Reference-Format}


\end{document}